\documentclass[preprint,nofootinbib,showpacs,preprintnumbers,amsmath,amssymb]{revtex4}

\usepackage{graphicx}
\usepackage{dcolumn}
\usepackage{bm}

\begin{document}

\title{On the Wave Aspect of Relativistic Quantum Fields}

\author{Minoru  Omote}
\email{omote@phys-h.keio.ac.jp}
\author{Susumu Kamefuchi}
\thanks{Professor emeritus,  University of Tsukuba, Ibaraki, Japan}
 \affiliation{Department of Physics, Keio University, Hiyoshi, Yokohama, Japan.}

\date{\today}

\begin{abstract}

In the state-vector space for relativistic quantum fields a new set of basis vectors are introduced, which are taken to be eigenstates of the field operators themselves. The corresponding eigenvalues are then interpreted as representing  matter waves associated with the respective quantum fields. The representation, based on such basis vectors, or the wave-representation naturally emphasizes the wave aspect of the system, in contrast with the usual, Fock or particle-representation  emphasizing the particle aspect. For the case of a relativistic, free neutral field, the wave-representation is explicitly constructed, and its mathematical properties as well as physical implications are studied in detail. It is expected that such an approach will find useful applications, e.g., in quantum optics.
\end{abstract}

\maketitle

\section{\label{sec:level1}Introduction}

In discussing various problems of particles and fields it is customary to base the formalism upon the Fock space in which the basis vectors are taken to be eigenstates of the partcle number operators \cite{FOCK}. Everything is expressed here in terms of those particles that arise from the quantum fields concerned. Thus, for example, the vacuum state is identified with the state in which no particles exist. Such a way of expressing quantum fields, or the Fock representation, is the one emphasizing the particle aspect thereof, hence may  be called the {\it particle-representation} of quantum fields. The particle aspect is thereby made diagonal, so to speak.

From the standpoint of the wave-particle duality of matter, however, we must admit that the above-mentioned approach is concerned only with a half of the story. That is to say, another half, i.e., the wave aspect of quantum fields is, to some extent, pushed to the background of the scene. Thus, invoking the transformation theory of quantum mechanics we can equally expect the existence of an alternative representation in which the wave aspect is made  diagonal, that is, the wave property comes to the foreground. Such a representation will hereafter be called the  {\it wave-representation}, and this should therefore be the one that is complementary to the  particle-representation. 

Now as a matter of course the wave property of quantum fields is most explicitly embodied by the field operators themselves, and their eigenvalues are to be understood as representing the corresponding matter waves. Thus, in the wave-representation we choose its basis vectors to be eigenstates of the field operators . In what follows a detailed study will be made of various problems concerning this representation.

For this purpose we consider, as an example, the simplest case of relativistic fields, that is, a free neutral scalar field $\phi(\vec{x},t)$, being given in the conventional notation as
\begin{equation}
\phi(\vec{x},t)=\sum_{\vec{k}}\sqrt{\frac{\hbar c^{2}}{2\omega_{k} L^{3}}}~\Bigl\{a(\vec{k})~e^{i\vec{k}\vec{x}-i\omega_{k}t}+a^{\dagger}(\vec{k})~e^{-i\vec{k}\vec{x}+i\omega_{k}t}\Bigr\},  
\label{eq:phi1}
\end{equation}
where $\hbar=$ (the Planck const.)/$2\pi$, $\omega_{k}\equiv c\sqrt{\vec{k}^{2}+\mu^{2}}, \mu=m c/\hbar$ and $k_{j}=(2\pi/L)n_{j}$ with $n_{j}=0,\pm 1,\pm 2,\cdots (j=1,2,3)$. The corresponding  momentum operator $\pi(\vec{x},t)$ is then written as
\begin{equation}
\pi(\vec{x},t)=-i \sum_{\vec{k}}\sqrt{\frac{\hbar \omega_{k}}{2c^{2} L^{3}}}~\Bigl\{a(\vec{k})~e^{i\vec{k}\vec{x}-i\omega_{k}t}-a^{\dagger}(\vec{k})~e^{-i\vec{k}\vec{x}+i\omega_{k}t}\Bigr\}.
\label{eq:pi1}
\end{equation}
Here, the operators $a(\vec{k})$ and $a^{\dagger}(\vec{k})$ are supposed to satisfy the usual commutation relations: $[a(\vec{k}),a^{\dagger}(\vec{k}^{\prime})]=\delta_{\vec{k},\vec{k}^{\prime}}$ and $[a(\vec{k}),~a(\vec{k}^{\prime})]=0$. Our problem  is then to solve the eigenvalue problems of the operators $\phi$ or $\pi$.

Before doing so we must make a few remarks, however. (a)  Since $[\phi(\vec{x},t),\phi(\vec{x}^{\prime},t^{\prime})]=[ \pi(\vec{x},t), \pi(\vec{x}^{\prime},t^{\prime})]=0$ does not hold true for $t \ne t^{\prime}$ in general, we must be content with solving the eigenvalue problems for equal-time quantities, e.g., $\phi(\vec{x},0)$'s or $\pi(\vec{x},0)$'s. Incidentally, such a situation may be contrasted with the case of non-relativistic fields, where the field commutativity for $t \ne t^{\prime}$ as well as $t=t^{\prime}$ holds true in most of the interesting cases: cf. for example \cite{KO}. (b) There are, in fact, two ways of dealing with our problem, depending on which space $-$ the $\vec{x}$- or $\vec{k}$- space $-$ is discretized. In what follows we adopt the latter way, as already done in (1) and (2). In this case the eigenvalues become some continuous functions in $\vec{x}$, to express  the spatial shapes of matter waves concerned . (c) As will be shown in sect 6, our wave-representation turns out mathematically to be inequivalent with the particle-representation. What we would like to do here, however, lies in   formulating both wave states $|{\rm wave}>$ and particle states $|{\rm particle}>$  within one and the same representation, so that the usual, complementary interpretation about the wave-particle duality is retained, i.e., $<{\rm wave}|{\rm particle}> \ne 0$ in general. Thus, we shall restrict our consideration beforehand to a finite region of the discretized $\vec{k}$- space. In this way we avoid the mathematical complications inherent in the system of infinite degrees of freedom.

In view of all this we shall hereafter consider, instead of  (\ref{eq:phi1}) and (\ref{eq:pi1}), the operators of the following, rearranged and restricted  forms: 
\begin{eqnarray}
&&\phi(\vec{x},0)=\sqrt{\frac{\hbar c^{2}}{2\omega_{0}L^{3}}}\Bigl\{ a(0)+a^{\dagger}(0)\Bigr\} \nonumber  \\ 
&&\hspace{1cm}+\sum_{\vec{k}\in K}\sqrt{\frac{\hbar c^{2}}{2\omega_{k} L^{3}}}~\Bigl\{(a(\vec{k})+a^{\dagger}(-\vec{k}))~e^{i\vec{k}\vec{x}}+\bigl(a(-\vec{k})+a^{\dagger}(\vec{k})\bigr)~e^{-i\vec{k}\vec{x}}\Bigr\}, \nonumber  \\
&&   \label{eq:phi2} \\
&&\pi(\vec{x},0)=-i\sqrt{\frac{\hbar\omega_{0}}{2c^{2}L^{3}}}\Bigl\{a(0)-a^{\dagger}(0)\Bigr\} \nonumber \\ 
&&\hspace{1cm}-i \sum_{\vec{k}\in K}\sqrt{\frac{\hbar \omega_{k}}{2c^{2} L^{3}}}~\Bigl\{(a(\vec{k})-a^{\dagger}(-\vec{k})~e^{i\vec{k}\vec{x}}+(a(-\vec{k})-a^{\dagger}(\vec{k})~e^{-i\vec{k}\vec{x}}\Bigr\} \label{eq:pi2}.
\end{eqnarray}
Here $K$ denotes some finite part of the region $\bar{K}$, whereas  $\bar{K}$ is defined as an arbitrarily chosen half of the $\vec{k}$-space, excepting $\vec{k}=0$, e.g., the region with $n_{3} \ge 0$, excepting $n_{1}=n_{2}=n_{3}=0$.

Now in these forms the commutability $[a(\vec{k}) \pm a^{\dagger}(-\vec{k}), a(\vec{k}^{\prime}) \pm a^{\dagger}(-\vec{k}^{\prime})]=0$ (under the usual double-sign convention) holds true for all $\vec{k}$ and $\vec{k}^{\prime}$, hence our eigenvalue problem is essentially reduced to those of the operators $\bigl(a(0) \pm a^{\dagger}(0)\bigr)$ and of $\bigl(a(\vec{k}) \pm a^{\dagger}(-\vec{k})\bigr)$ with $\vec{k} \in K$. Then, simultaneous eigenstates of all $\phi(\vec{x},0)$'s or of $\pi(\vec{x},0)$'s are simply given as direct products of eigenstates  of the above-mentiond operators.

In view of this we begin in sect 2 and sect 3 with studying one- and two-mode systems described, respectively, by one set and two sets of creation and annihilation operators. Here the required eigenstates are explicitly given in terms of the basic operators $a(\vec{k})$ and $a^{\dagger}(\vec{k})$. In sect 4 we turn our attention to some mathematical relations arising from our problem,  by-products being various formulae to relate Hermite to Laguerre polynomials.  Sect 5 is devoted to  eigenvalue problems of field operators $\phi(\vec{x},0)$ and $\pi(\vec{x},0)$, and the wave-representation is  constucted explicitly.  

On the basis of the results in the preceding sections we introduce in sect 6 a new kind of vacuum state, or the {\it wave-vacuum} as the ground state of the wave-representation, and examine how this vacuum differs from, and relates to, the usual vacuum in the particle-representation, or the {\it particle-vacuum}. Lastly in sect 7 two further remarks are added. The first is concerned with two different meanings which the word {\it superposition} has in quantum field theory, and the second with applications of our wave-representaion to the cases of the electromagnetic and other fields. Details of mathematical proofs are all relegated to Appendix {\bf 1}$\sim$ {\bf 5}.

\section{State vectors of one - and two -mode systems}

In this section we study some basic properties of state vectors for one- and two-mode systems, described respectively by $a$ and $a^{\dagger}$ and by $a,a^{\dagger},b$ and $b^{\dagger}$, where $[a,a^{\dagger}]=1, [b,b^{\dagger}]=1$ and $[a,b]=[a,b^{\dagger}]=0$. In so doing we shall base our arguments exclusively on the algebraic properties of these operators.

\subsection{Case of one-mode systems}
We begin by considering a one-mode system described by $a$ and $a^{\dagger}$ and re-derive some of the well-known results purely in an algebraic manner. First   we note that the eigenstates of the operator $(a+\epsilon a^{\dagger})$ with eigenvalue $\xi$ are given in the form
\begin{equation}
|\xi;\epsilon>=\frac{1}{(2\pi)^{1/4}}~e^{-|\xi|^{2}/4} ~e^{-\epsilon (a^{\dagger})^{2}/2}~e^{\xi a^{\dagger}}|0>,  
\label{eq:state1}
\end{equation}
where $\epsilon=\pm, |0>$ is the state satisfying $a|0>=0$ and $<0|0>=1$. Here we see also $\xi^{\ast}=\epsilon \xi$, so that for $\epsilon=+(-)$ all eigenvalues $\xi$ are real ( pure imaginary): $\xi=\xi^{\ast}=\xi_{r} ~(\xi=-\xi^{\ast}=i\xi_{i}$). This should be so, because the above operator with $\epsilon=+ (-)$ equals essentially $\sqrt{2}q~ (\sqrt{2}ip)$ in the usual notation for canonical variables. In fact, using (\ref{eq:state1}) we can easily confirm that
\begin{equation}
(a+\epsilon a^{\dagger})|\xi;\epsilon> =\xi |\xi;\epsilon>,~~~<\xi;\epsilon|(a+\epsilon a^{\dagger})=<\xi;\epsilon|\epsilon \xi^{\ast}.
\label{eq:eigeneq1}
\end{equation}
To do this we have only to note that $\exp[\epsilon (a^{\dagger})^{2}/2] (a+\epsilon a^{\dagger})\exp[-\epsilon (a^{\dagger})^{2}/2]=a$. As shown in Appendix {\bf 1} we see also that the following relations hold true:
\begin{eqnarray}
&&<\xi_{r}^{\prime};+|\xi_{r};+>=\delta(\xi_{r}^{\prime}-\xi_{r}),~~~<\xi_{i}^{\prime};-|\xi_{i};->=\delta(\xi_{i}^{\prime}-\xi_{i}),  \label{eq:ortho1}  \\
&&<\xi_{i}^{\prime};-|\xi_{r};+>=\frac{1}{2\sqrt{\pi}}~e^{-i\xi_{i}^{\prime }\xi_{r}/2},  
\label{eq:trans1}  \\
&&\int_{-\infty}^{\infty} d \xi_{r} ~|\xi_{r};+><\xi_{r};+|={\rm I},~~\int_{-\infty}^{\infty} d \xi_{i} ~|\xi_{i};-><\xi_{i};-|={\rm I}.
\label{eq:complete1}
\end{eqnarray}

On the other hand, the eigenstates $|n> $ of the operator $a^{\dagger}a$ with eigenvalue $n( =0,1,2,\cdots)$ are given as usual by
\begin{equation}
|n>\equiv \frac{1}{\sqrt{n!}}(a^{\dagger})^{n}|0>,
\label{eq:nstate1}
\end{equation}
and satisfy
\begin{eqnarray}
&& a^{\dagger}a |n>=n|n>, ~~~~~<n^{\prime}|n>=\delta_{nn^{\prime}},   \nonumber \\
&& \sum_{n} |n><n|={\rm I}.
\label{eq:nstate2}
\end{eqnarray}
As proved in Appendix {\bf 2}, we then have
\begin{eqnarray}
&&<n|\xi_{r};+>=\sqrt{\frac{1}{\sqrt{2\pi}n!}}~ e^{-\xi_{r}^{2}/4}~H_{n}(\xi_{r}),  \label{eq:H1} \\
&&<n|i\xi_{i};->=i^{n} \sqrt{\frac{1}{\sqrt{2\pi}n!}}~ e^{-\xi_{i}^{2}/4}~H_{n}(\xi_{i}), \label{eq:H2}
\end{eqnarray}
where $H_{n}(x)$'s are Hermite polynomials with $n=0,1,2,\cdots$.

From (\ref{eq:H1}) and (\ref{eq:H2}) we obtain at once
\begin{eqnarray}
&&|\xi_{r};+>=\frac{1}{(2\pi)^{1/4}}~e^{-\frac{\xi_{r}^{2}}{4}}\sum_{n=0}^{\infty} \frac{1}{\sqrt{n!}}H_{n}(\xi_{r}) |n>,    \label{eq:state1-1}  \\
&&|i\xi_{i};->=\frac{1}{(2\pi)^{1/4}}~e^{-\frac{\xi_{i}^{2}}{4}}\sum_{n=0}^{\infty} \frac{i^{n}}{\sqrt{n!}}H_{n}(\xi_{i}) |n>, \label{eq:state1-2}
\end{eqnarray}
which for the cases $\xi_{r}=0$  or $\xi_{i}=0$ result in
\begin{eqnarray}
&&|0;+>=\frac{1}{(2\pi)^{1/4}}~\sum_{n=0}^{\infty} \frac{(-1)^{n}(2n-1)!!}{\sqrt{(2n)!}}~~|2n>,    \label{eq:state1-3}  \\
&&|0;->=\frac{1}{(2\pi)^{1/4}}~\sum_{n=0}^{\infty} \frac{(2n-1)!!}{\sqrt{(2n)!}} ~~|2n>. \label{eq:state1-4}
\end{eqnarray}
The converse of the  relation (\ref{eq:state1-1}), for example, turns out to be 
\begin{equation}
|n>=\sqrt{\frac{1}{\sqrt{2\pi}n!}}\int_{-\infty}^{\infty} d \xi_{r} ~e^{-\xi_{r}^{2}/4} H_{n}(\xi_{r})~|\xi_{r};+>. \label{eq:nstate-1}
\end{equation}
Thus for $n=0$, in particular, the above relation becomes
\begin{equation}
|0>=\bigl(\frac{1}{2\pi}\Bigr)^{1/4} \int_{-\infty}^{\infty} d \xi_{r} ~e^{-\xi_{r}^{2}/4}~|\xi_{r};+>, \label{eq:0state1}
\end{equation}
which corresponds to a Gaussian wave-packet made up of the  $|\xi_{r};+>$'s.

\subsection{Case of two-mode systems}
Let us now turn to the case of a two-mode system. State vectors for such systems can, of course, be given as the direct product of those of type (\ref{eq:state1}) for the respective one-mode systems. In fact, the state
\begin{equation}
|\xi, \eta;\epsilon,\epsilon^{\prime}>=|\xi;\epsilon>\otimes ~|\eta;\epsilon^{\prime}>=\frac{1}{\sqrt{2\pi}}e^{-(|\xi|^{2}+|\eta|^{2})/4}~e^{-\{\epsilon(a^{\dagger})^{2}+\epsilon^{\prime}(b^{\dagger})^{2}\}/2}e^{\xi a^{\dagger}+\eta b^{\dagger}}|0>
\label{eq:state2}
\end{equation}
with $|0>$ being such that $a|0>=b|0>=0$ and $<0|0>=1$, provides a simultaneous eigenstate of the operator $(a+\epsilon a^{\dagger}) $ and $(b+\epsilon^{\prime}b^{\dagger})$, satisfying
\begin{eqnarray}
&&(a+\epsilon a^{\dagger})|\xi,\eta;\epsilon,\epsilon^{\prime}>=\xi|\xi,\eta;\epsilon,\epsilon^{\prime}>,~~~(b+\epsilon^{\prime} b^{\dagger})|\xi,\eta;\epsilon,\epsilon^{\prime}>=\eta|\xi,\eta;\epsilon,\epsilon^{\prime}>,   \label{eq:eigeneq2}\\
&&<\xi,\eta;\epsilon,\epsilon^{\prime}|(a+\epsilon a^{\dagger})=<\xi,\eta;\epsilon,\epsilon^{\prime}|\epsilon \xi^{\ast},~~~<\xi,\eta;\epsilon,\epsilon^{\prime}|(b+\epsilon^{\prime} b^{\dagger})=<\xi,\eta;\epsilon,\epsilon^{\prime}|\epsilon^{\prime} \eta^{\ast},   \nonumber  \\
&& \label{eq:eigeneqq3}
\end{eqnarray}
where $\epsilon, \epsilon^{\prime}=\pm$.

\section{Another type of basis vectors for two-mode systems}
It is  more convenient for our present purpose, however, to employ another type of simultaneous eigenstates $|\xi, \epsilon \xi^{\ast}>$, defined by
\begin{equation}
|\xi,\epsilon \xi^{\ast}> \equiv \frac{1}{\sqrt{\pi}}e^{-|\xi|^{2}/2}~e^{-\epsilon a^{\dagger}b^{\dagger}}~e^{\xi a^{\dagger}+\epsilon \xi^{\ast} b^{\dagger}}|0>.
\label{eq:state3}
\end{equation}
That this is a simultaneous eigenstate of $(a+\epsilon b^{\dagger})$ and $(b+\epsilon a^{\dagger})$ can be seen in a way similar to the case of (\ref{eq:state1}):
\begin{eqnarray}
&&(a+\epsilon b^{\dagger})|\xi, \epsilon \xi^{\ast}>=\xi |\xi,\epsilon \xi^{\ast}>,  ~~~(b+\epsilon a^{\dagger})|\xi, \epsilon \xi^{\ast}>=\epsilon \xi^{\ast} |\xi,\epsilon \xi^{\ast}>,
\label{eq:state3-1}  \\
&&<\xi,\epsilon \xi^{\ast}|(a+\epsilon b^{\dagger})=<\xi,\epsilon \xi^{\ast}|\xi, ~~~<\xi,\epsilon \xi^{\ast}|(b+\epsilon a^{\dagger})=<\xi,\epsilon \xi^{\ast}|\epsilon \xi^{\ast}. \label{eq:state3-2}
\end{eqnarray}
Unlike the previous case  the eigenvalues $\xi$ and $\epsilon \xi^{\ast}$ here are complex numbers in general.

For the  state-vectors (\ref{eq:state3}) the following relations are then shown to be true:
\begin{eqnarray}
&&<\xi^{\prime},\epsilon \xi^{\prime \ast}|\xi,\epsilon \xi^{\ast}>=\delta(\xi_{r}-\xi_{r}^{\prime})\delta(\xi_{i}-\xi_{i}^{\prime}), \label{eq:ortho3} \\
&&<\xi^{\prime},-\xi^{\prime \ast}|\xi,\xi^{\ast}>=\frac{1}{2\pi}\exp\Bigl(\frac{\xi^{\prime \ast}\xi-\xi^{\prime}\xi^{\ast}}{2}\Bigr)=\frac{1}{2\pi}\exp\Bigl(i(\xi_{r}^{\prime}\xi_{i}-\xi_{i}^{\prime}\xi_{r})\Bigr),
\label{eq:transf3}
\end{eqnarray}
where we have put $\xi=\xi_{r}+i\xi_{i}$, and similarly for $\xi^{\prime}$. Proofs of these relations are given in Appendix {\bf 3}.

On the other hand, simultaneous eigenstates of the operators $a^{\dagger}a$ and $b^{\dagger}b$, with  eigenvalues being $n$ and $m$ respectively, are given by
\begin{equation}
|n,m>\equiv \frac{1}{\sqrt{n!m!}}~(a^{\dagger})^{n}(b^{\dagger})^{m}|0>,
\label{eq:nmstate}
\end{equation}
and satisfy
\begin{eqnarray}
&& a^{\dagger}a |n,m>=n |n,m>,~~b^{\dagger}b |n,m>=m |m>,  \label{eq:eigeneqnm} \\
&& <n^{\prime},m^{\prime}|n,m>=\delta_{nn^{\prime}}\delta_{mm^{\prime}},~~\sum_{n,m} |n,m><n,m|={\rm I}.  \label{eq:orthocomplete}
\end{eqnarray}

As shown in Appendix {\bf 4} the following relations then hold true:
\begin{eqnarray}
&&<n,m|\xi,\epsilon \xi^{\ast}>=\frac{1}{\sqrt{\pi}}\sqrt{\frac{m!}{ n!}}e^{-|\xi|^{2}/2}\xi^{n-m}(-\epsilon)^{m}  L_{m}^{n-m}(|\xi|^{2})  ~~~ {\rm for} ~~n-m\ge 0, \nonumber  \\
&&<n,m|\xi,\epsilon \xi^{\ast}>=\frac{1}{\sqrt{\pi}}\sqrt{\frac{n!}{ m!}}e^{-|\xi|^{2}/2} (\epsilon \xi^{\ast})^{m-n}(-\epsilon)^{n}  L_{n}^{m-n}(|\xi|^{2})  ~~~ {\rm for} ~~m-n \ge 0,  \nonumber \\
&&\label{eq:nm}
\end{eqnarray}
where $L_{n}^{m}(x)$'s denote associated Laguerre polynomials, defined by
\begin{equation}
L_{n}^{m}(x) \equiv \frac{1}{n!}e^{x} x^{-m} \frac{d^{n}}{d x^{n}}\bigl(e^{-x} x^{n+m}\bigr)=\sum_{r=0}^{n} (-1)^{r} ~_{n+m}C_{n-r} ~\frac{x^{r}}{r!}.
\label{eq:Laguerre1}
\end{equation}
Incidentally , the appearance of Laguerre polynomials in certain Fock-space problems has been known for some time: see for example \cite{FS,F,FF}.

Using (\ref{eq:nm}) and the last relation in (\ref{eq:orthocomplete}) we can express $|\xi, \epsilon \xi^{\ast}>$ in terms of $|n,m>$:
\begin{eqnarray}
|\xi,\epsilon \xi^{\ast}>&=&\frac{e^{-|\xi|^{2}/2}}{\sqrt{\pi}}\Bigl\{\sum_{n=0}^{\infty}\sum_{m=0}^{n} \sqrt{\frac{m!}{n!}}\xi^{n-m} (-\epsilon)^{m} L_{m}^{n-m}(|\xi|^{2})~|n,m>  \nonumber  \\
&&\hspace{2cm} +\sum_{m=1}^{\infty}\sum_{n=0}^{m-1} \sqrt{\frac{n!}{m!}}(\epsilon \xi^{\ast})^{m-n} (-\epsilon)^{n} L_{n}^{m-n}(|\xi|^{2})~|n,m>\Bigr\} . \nonumber  \\
\label{eq:state3-4}
\end{eqnarray}
A straightforward calculation by use of (\ref{eq:state3-4}) leads us then to the completeness relation:
\begin{equation}
\int _{-\infty}^{\infty} d^{2}\xi ~|\xi,\epsilon \xi^{\ast}><\xi,\epsilon \xi^{\ast}|={\rm I},\label{eq:complete3}
\end{equation}
where $\int d^{2} \xi \equiv \int_{-\infty}^{\infty} d\xi_{r} \int_{-\infty}^{\infty} d\xi_{i}$. Calculational details are given in Appendix {\bf 5}. The relation (\ref{eq:complete3}) together with (\ref{eq:ortho3}) shows that the basis vectors $|\xi,\epsilon \xi^{\ast}>$'s form a complete ortho-{\it normal} system.

The converse relations of (\ref{eq:state3-4}) can also be obtained from (\ref{eq:nm}) and (\ref{eq:complete3});
\begin{eqnarray}
&&|n,m>=\frac{1}{\sqrt{\pi}}\sqrt{\frac{m!}{ n!}}~\int d^{2}\xi e^{-|\xi|^{2}/2}\xi^{n-m}(-\epsilon)^{m}  L_{m}^{n-m}(|\xi|^{2}) ~|\xi,\epsilon \xi^{\ast}> ~~~ {\rm for} ~~n-m\ge 0, \nonumber  \\
&&|n,m>=\frac{1}{\sqrt{\pi}}\sqrt{\frac{n!}{ m!}}~\int d^{2} \xi e^{-|\xi|^{2}/2} (\epsilon \xi^{\ast})^{m-n}(-\epsilon)^{n}  L_{n}^{m-n}(|\xi|^{2})|\xi,\epsilon\xi^{\ast}>  ~~~ {\rm for} ~~m-n \ge 0 .\label{eq:nm-1}  \nonumber  \\
&&
\end{eqnarray}
Similarly, by use of  relation (\ref{eq:transf3}) we can express, e.g., $|\xi,\xi^{\ast}>$ in terms of $|\xi,-\xi^{\ast}>$'s:
\begin{equation}
|\xi, \xi^{\ast}>=\frac{1}{2\pi}\int d \xi_{r}^{\prime} d \xi_{i}^{\prime}  e^{(\xi^{\prime \ast}\xi-\xi^{\prime}\xi^{\ast})/2}|\xi^{\prime},-\xi^{\prime \ast}>. \label{eq:trans3-1}
\end{equation}
Thus for a state $|~>=\int d^{2}\xi \psi(\xi)~|\xi,\xi^{\ast}>$ with $\psi(\xi)\equiv <\xi,\xi^{\ast}|~>$ we find by virtue of (\ref{eq:trans3-1}) that
\begin{eqnarray}
&& (a+b^{\dagger})|~>=\int d^{2}\xi ~\xi ~\psi(\xi)~|\xi,\xi^{\ast}>, \label{eq:ab-1} \\
&& (a-b^{\dagger}) |~>=\int d^{2}\xi \Bigl[\Bigl(\frac{\partial}{\partial \xi_{r}}+i\frac{\partial}{\partial \xi_{i}}\Bigr)\psi(\xi)\Bigr]|\xi,\xi^{\ast}>.
\label{eq:ab}
\end{eqnarray}

By putting $n=m=0$ in (\ref{eq:nm-1}) and $\xi=\xi^{\ast}=0$ in (\ref{eq:state3-4}) we obtain respectively
\begin{equation}
|n=0,m=0>=\frac{1}{\sqrt{\pi}}\int d^{2}\xi ~e^{-|\xi|^{2}/2}~|\xi,\xi^{\ast}>,
\label{eq:vacuum2}
\end{equation}
and
\begin{equation}
|\xi=0,\xi^{\ast}=0>=\sum_{n=0}^{\infty} \frac{(-1)^{n}}{\sqrt{\pi }}~|n,n>.
\label{eq:vacuum3}
\end{equation}
The norm of (\ref{eq:vacuum3}) is diverging because it is {\it normalized} according to  (\ref{eq:ortho3}). The above relation, (\ref{eq:vacuum2}) and (\ref{eq:vacuum3}), exihibit the complementary nature between the description using $|n,m>$'s and the one using $|\xi,\xi^{\ast}>$'s.

\section{Mathematical aspects of the problem}

In this section we make a digression to give a few remarks of mathematical nature. Some interesting mathematical formulae are also derived as by-products.

\subsection{Eigenstates of the operator $a^{\dagger}a$}
Let us first consider the matrix element of the operator $a^{\dagger}a$ with respect to two states of type (\ref{eq:state1}) with $\epsilon=+$ $~(\xi=\xi_{r}, \xi^{\prime}=\xi^{\prime}_{r}$):
\begin{eqnarray}
<\xi^{\prime};+|a^{\dagger}a|\xi;+>&=&\lim_{d \to 0} \frac{1}{2\sqrt{\pi d}}\Bigl[\Bigl(-\frac{1}{2}+\frac{\xi^{2}}{4}+\frac{1}{2d}-\frac{(\xi-\xi^{\prime})^{2}}{4d^{2}}\Bigr)e^{-(\xi-\xi^{\prime})^{2}/4d}\Bigr]  \nonumber  \\
&&=\Bigl(-\frac{1}{2}+\frac{\xi^{2}}{4}-\frac{\partial^{2}}{\partial \xi^{2}}\Bigr)\delta(\xi-\xi^{\prime}),
\label{eq:number1}
\end{eqnarray}
where we have made use of a technique as mentioned in Appendix {\bf 3}. By virtue of (\ref{eq:number1}) the eigenstate $|\Psi_{n}>$ of the number operator $a^{\dagger}a$ with eigenvalue $n$ can be written as
\begin{equation}
\Bigl(-\frac{1}{2}+\frac{\xi^{2}}{4}-\frac{\partial^{2}}{\partial \xi^{2}}\Bigr)<\xi;+|\Psi_{n}>=n<\xi;+|\Psi_{n}>.
\label{eq:number2}
\end{equation}
Assuming an exponential damping for $|\xi| \to \infty$ and putting $<\xi;+|\Psi_{n}>\equiv e^{-\xi^{2}/4}f_{n}(\xi)$, we find from (\ref{eq:number2}) that
\begin{equation}
\frac{d^{2}}{d \xi^{2}}f_{n}(\xi)-\xi \frac{d}{d \xi}f_{n}(\xi)+n f_{n}(\xi)=0.
\label{eq:number3}
\end{equation}
As is well known, the solution of (\ref{eq:number3}) is given in terms of Hermite polynomials $H_{n}(\xi)$ with eigenvalue $n=0,1,2,\cdots$, so that $|\Psi_{n}>$ agrees with $|n>$ of (\ref{eq:nstate-1}). It is worth emphasizing here that the interpretation of $a^{\dagger}a$ as the number operator is made possible only under the above-mentioned boundary condition (another, exponentially increasing boundary condition is to be physically excluded).

Let us next rewrite the left-hand side of (\ref{eq:number1}) by use of (\ref{eq:state1-1} ) as follows:
\begin{eqnarray}
<\xi^{\prime};+|a^{\dagger}a|\xi;+>&=&\sum_{n}~ n <\xi^{\prime};+|n><n|\xi;+>  \nonumber  \\
&=&\frac{1}{\sqrt{2\pi}}\sum_{n} ~\frac{n}{n!}~e^{-(\xi^{2}+\xi^{\prime 2})/4}~H_{n}(\xi^{\prime})H_{n}(\xi).
\label{eq:Hn1}
\end{eqnarray}
Combining this with (\ref{eq:number1}) we obtain a mathematical formula for real $x$ and $y$ such that
\begin{equation}
\frac{1}{\sqrt{2\pi}}\sum_{n} ~\frac{n}{n!}~e^{-(x^{2}+y^{ 2})/4}~H_{n}(x)H_{n}(y)=\Bigl(-\frac{1}{2}+\frac{x^{2}}{4}-\frac{\partial^{2}}{\partial x^{2}}\Bigr)\delta(x-y).
\label{eq:Hn2}
\end{equation}
It is also easy to generalize this to
\begin{equation}
\frac{1}{\sqrt{2\pi}}\sum_{n} ~\frac{P(n)}{n!}~e^{-(x^{2}+y^{ 2})/4}~H_{n}(x)H_{n}(y)=P\Bigl(-\frac{1}{2}+\frac{x^{2}}{4}-\frac{\partial^{2}}{\partial x^{2}}\Bigr)~\delta(x-y),
\end{equation}
where $P(n)$ is an arbitrary polynomial in $n$.

\subsection{Simultaneous eigenstates of the operators $a^{\dagger}a$ and $b^{\dagger}b$}

Performing an algebraic calculation similar to, but more lengthy than, the above we find the matrix elements of the operators $a^{\dagger}a$ and $b^{\dagger}b$ with respect to two states of type (\ref{eq:state3}) with $\epsilon = +$ to be
\begin{eqnarray}
&&<\xi,\xi^{ \ast}|a^{\dagger}a|\xi^{\prime},\xi^{\prime \ast}>=\Bigl[-\frac{1}{2}+\frac{|\xi^{\prime}|^{ 2}}{4}+\frac{i}{2}\Bigl(\xi_{i}\frac{\partial}{\partial \xi_{r}^{\prime}}-\xi_{r}\frac{\partial}{\partial \xi_{i}^{\prime}}\Bigr)  \nonumber  \\
&&\hspace{4cm}-\frac{1}{4}\Bigl(\frac{\partial^{2}}{\partial \xi_{r}^{\prime 2}}+\frac{\partial^{2}}{\partial \xi_{i}^{\prime 2}}\Bigr)\Bigr]\delta(\xi_{r}-\xi_{r}^{\prime})\delta(\xi_{i}-\xi_{i}^{\prime}),  \label{eq:number4} \\
&&<\xi,\xi^{ \ast}|b^{\dagger}b|\xi^{\prime},\xi^{\prime \ast}> =\Bigl[-\frac{1}{2}+\frac{|\xi^{\prime}|^{ 2}}{4}-\frac{i}{2}\Bigl(\xi_{i}\frac{\partial}{\partial \xi_{r}^{\prime}}-\xi_{r}\frac{\partial}{\partial \xi_{i}^{\prime}}\Bigr) \nonumber \\
&&\hspace{4cm}-\frac{1}{4}\Bigl(\frac{\partial^{2}}{\partial \xi_{r}^{\prime 2}}+\frac{\partial^{2}}{\partial \xi_{i}^{\prime 2}}\Bigr)\Bigr]\delta(\xi_{r}-\xi_{r}^{\prime})\delta(\xi_{i}-\xi_{i}^{\prime}). \label{eq:number5}
\end{eqnarray}

When use is made of (\ref{eq:number4}) and (\ref{eq:number5}) the simultaneous eigennstate $|\Psi_{nm}>$ of the operator $a^{\dagger}a$ and $b^{\dagger}b$, with eigenvalues being $n$ and $m$ respectively, can be written as
\begin{eqnarray}
&&\Bigl[-\frac{1}{2}+\frac{|\xi|^{ 2}}{4}-\frac{i}{2}\Bigl(\xi_{i}\frac{\partial}{\partial \xi_{r}}-\xi_{r}\frac{\partial}{\partial \xi_{i}}\Bigr)-\frac{1}{4}\Bigl(\frac{\partial^{2}}{\partial \xi_{r}^{ 2}}+\frac{\partial^{2}}{\partial \xi_{i}^{ 2}}\Bigr)\Bigr]<\xi,\xi^{\ast}|\Psi_{nm}> \nonumber  \\
&&\hspace{7cm}=n<\xi,\xi^{\ast}|\Psi_{nm}>, \label{eq:number4-1} \\
&&\Bigl[-\frac{1}{2}+\frac{|\xi|^{ 2}}{4}+\frac{i}{2}\Bigl(\xi_{i}\frac{\partial}{\partial \xi_{r}}-\xi_{r}\frac{\partial}{\partial \xi_{i}}\Bigr)-\frac{1}{4}\Bigl(\frac{\partial^{2}}{\partial \xi_{r}^{ 2}}+\frac{\partial^{2}}{\partial \xi_{i}^{ 2}}\Bigr)\Bigr]<\xi,\xi^{\ast}|\Psi_{nm}> \nonumber \\
&&\hspace{7cm}=m<\xi,\xi^{\ast}|\Psi_{nm}>. \label{eq:number5-1}
\end{eqnarray}
Assuming again an exponential damping for $|\xi|\to \infty$ and putting $<\xi,\xi^{\ast}|\Psi_{nm}>=(\xi^{\ast})^{\alpha} e^{-|\xi|^{2}/2} f_{nm}(|\xi|^{2})$ we obtain  from (\ref{eq:number4-1}) and (\ref{eq:number5-1}):
\begin{eqnarray}
&& |\xi|^{2} ~f_{nm}^{\prime\prime}(|\xi|^{2})+(\alpha+1-|\xi|^{2})~f_{nm}^{\prime}(|\xi|^{2}+(n-\alpha)~f_{nm}(|\xi|^{2})=0,  \nonumber \\
&&|\xi|^{2} ~f_{nm}^{\prime\prime}(|\xi|^{2})+(\alpha+1-|\xi|^{2})~f_{nm}^{\prime}(|\xi|^{2}+m~f_{nm}(|\xi|^{2})=0,\label{eq:number6}
\end{eqnarray}
where the primes attached to $f$ mean the derivative with respect to $|\xi|^{2}$. For the above two equations to be compatible we must have $\alpha=n-m$, and the solution of (\ref{eq:number6}) is $f_{nm}(|\xi|^{2}) \propto L_{m}^{n-m}(|\xi|^{2})$ for $n-m \ge 0$, thus agreeing with (\ref{eq:nm}). On the other hand, if we put $<\xi,\xi^{\ast}|\Psi_{nm}>=(\xi)^{\alpha} e^{-|\xi|^{2}/2}~g_{nm}(|\xi|^{2})$, we then find from (\ref{eq:number4-1}) and (\ref{eq:number5-1}):
\begin{eqnarray}
&& |\xi|^{2} ~g_{nm}^{\prime\prime}(|\xi|^{2})+(\alpha+1-|\xi|^{2})~g_{nm}^{\prime}(|\xi|^{2})+n~g_{nm}(|\xi|^{2})=0,  \\
&&|\xi|^{2} ~g_{nm}^{\prime\prime}(|\xi|^{2})+(\alpha+1-|\xi|^{2})~g_{nm}^{\prime}(|\xi|^{2})+(m-\alpha)~g_{nm}(|\xi|^{2})=0,
\end{eqnarray}
thereby implying that $\alpha=n-m$ and $g_{nm}(|\xi|^{2}) \propto L_{n}^{m-n}(|\xi|^{2}$ for $m-n \ge 0$ in agreement with (\ref{eq:nm}). Again we find that the above boundary condition for $|\xi| \to \infty$ underlies the discrete eigenvalues $0,1,2,\cdots$, required of the operators $a^{\dagger}a$ and $b^{\dagger}b$.

\subsection{Relations between Hermite and Laguerre polynomials}
Let us first evaluate the transformation function $<\xi_{1},\eta_{1};\epsilon_{1},\epsilon_{2}|\xi,\epsilon\xi^{\ast}>$ directly, that is, by subsituting the explicit expressions (\ref{eq:state2}) and (\ref{eq:state3}) therein and using the same technique as given in Appendix {\bf 1}. The result then is  as follows:
\begin{eqnarray}
&&<\xi_{1},\eta_{1};\epsilon_{1},\epsilon_{2}|\xi, \epsilon \xi^{\ast}> =\frac{1}{\sqrt{2}\pi}~e^{-(|\xi_{1}|^{2}+|\eta_{1}|^{2})/4}~e^{-|\xi|^{2}/2}~ \nonumber  \\
&&\hspace{4cm} \times <0|e^{\xi_{1}^{\ast} a+\eta_{1}^{\ast}b} e^{-\epsilon_{1} a^{2}/2}e^{-\epsilon_{2} b^{2}/2} e^{-\epsilon a^{\dagger}b^{\dagger}} e^{\xi a^{\dagger}+\epsilon \xi^{\ast} b^{\dagger}}|0> \nonumber  \\
&&\hspace{3.5cm}=\frac{1}{\sqrt{2}\pi}~e^{-(|\xi_{1}|^{2}+|\eta_{1}|^{2})/4}~e^{-|\xi|^{2}/2}~ I,
\label{eq:HL1}
\end{eqnarray}
where 
\begin{eqnarray}
&&\log  I \equiv \lim_{\lambda \to 1/2}\Bigl[\frac{1}{1-4\lambda^{2}\epsilon_{1}\epsilon_{2}}\Bigl\{-\epsilon_{2}\lambda (\xi_{1}^{\ast})^{2}-\epsilon_{1}\lambda (\eta_{1}^{\ast})^{2}-\epsilon \xi_{1}^{\ast}\eta_{1}^{\ast}+\xi_{1}^{\ast}(\xi+\epsilon_{2}\xi^{\ast}) \nonumber  \\
&&\hspace{3cm}+ \epsilon (1+2\epsilon_{1}\epsilon_{2}\lambda-4\epsilon_{1}\epsilon_{2}\lambda^{2})\eta_{1}^{\ast}\xi^{\ast}+2\epsilon \epsilon_{1}\lambda \eta_{1}^{\ast}\xi-2\epsilon_{1}\epsilon_{2}\lambda \xi \xi^{\ast}  \nonumber  \\
&& \hspace{3cm}-\epsilon_{1}\lambda \xi^{2}-\frac{\epsilon_{2}+2\epsilon_{1}\lambda-4\epsilon_{1}\lambda^{2}}{2}(\xi^{\ast})^{2}\Bigr\}-\frac{1}{2}\log(1-4\epsilon_{1}\epsilon_{2}\lambda^{2}) \Bigr]. \nonumber  \\
&&\hspace{4.5cm}
\label{eq:HL2}
\end{eqnarray}
The expression (\ref{eq:HL2}) is simplified, however, for the case $\epsilon_{1}=\epsilon_{2}=+$
\begin{equation}
I=\sqrt{2\pi}\delta(x+\epsilon y-2\xi_{r})~\exp\Bigl[\Bigl(\frac{(x^{2}+y^{2})}{2}+|\xi|^{2}+i\xi_{i}(x-\epsilon y)\Bigr)/2\Bigr],
\label{eq:HL3}
\end{equation}
where we have put $\xi_{1}=x,~\eta_{1}=y$. Combining (\ref{eq:HL1}) with (\ref{eq:HL3}) we obtain
\begin{equation}
<x,y;+.+|\xi,  \epsilon \xi^{\ast}> =\frac{1}{\sqrt{\pi}}~\delta(x+\epsilon y-2\xi_{r})~~e^{i\xi_{i}(x-\epsilon y)/2}.
\label{eq:HL4}
\end{equation}
Similarly, we have for the case $\epsilon_{1}=\epsilon_{2}=-$
\begin{equation}
<ix,iy;-.-|\xi,  \epsilon \xi^{\ast}> =\frac{1}{\sqrt{\pi}}~\delta \bigl(x-\epsilon y+2\xi_{i}\bigr)~~e^{-i\xi_{r}(x+\epsilon y)/2},
\label{eq:HL5}
\end{equation}
where $\xi_{1}\equiv i x, \eta_{1}\equiv i y$, and for the case $\epsilon_{1}=+, \epsilon_{2}=-$;
\begin{equation}
<x,iy;+,-|\xi,\epsilon \xi^{\ast}>=\frac{1}{2\pi}e^{i\epsilon xy/2+i\xi_{i}x-i\epsilon \xi_{r} y-i\xi_{r}\xi_{i}},
\label{eq:HL6}
\end{equation}
where  $\xi_{1}\equiv x, \eta_{1}\equiv i y$.

Next, by use of (\ref{eq:complete1}) and (\ref{eq:state1-1}) we can rewrite $<n.m|\xi,\epsilon\xi^{\ast}>$ with, e.g., $n-m\ge0$ as follows:
\begin{eqnarray}
&&<n,m|\xi,\epsilon \xi^{\ast}>= \int_{-\infty}^{\infty} d x d y~ <n,m|x,y;+,+><x,y:+,+|\xi,\epsilon \xi^{\ast}>  \nonumber  \\
&&\hspace{1cm}= \frac{1}{\sqrt{2\pi n!m!}}\int_{-\infty}^{\infty} d x d y~ <x,y:+,+|\xi,\epsilon \xi^{\ast}>e^{-\frac{x^{2}+y^{2}}{4}} H_{n}(x)~H_{m}(y).\nonumber \\
\label{eq:HL7}
\end{eqnarray}
Taking account of (\ref{eq:nm}) and (\ref{eq:HL4}) in the above, we obtain a formula to relate Hermite to Laguerre polynomials:
\begin{eqnarray}
&&\int_{-\infty}^{\infty}  d y~ e^{i\xi_{i}y}~e^{-y^{2}/2} ~H_{n}(y+\xi_{r})~H_{m}(y-\xi_{r})=\sqrt{2\pi}~m!~e^{-|\xi|^{2}}~\xi^{n-m}~L_{m}^{n-m}(|\xi|^{2})   \nonumber \\
&&\hspace{9cm} (n-m\ge 0) .
\label{eq:HL8}
\end{eqnarray}
Incidentally, if in (\ref{eq:HL7}) $|x,y;-,->$'s are employed instead of $|x,y;+,+>$'s, then the resulting formula will be essentially the same as (\ref{eq:HL8}). We add here that some special cases of (\ref{eq:HL8}) are found, for example, in \cite{HL}. 

If we put $\xi_{r}=0$ and write $\xi_{i}=x$ in (\ref{eq:HL8}), the formula will become
\begin{equation}
\int_{-\infty}^{\infty} dy~e^{ixy}~e^{-y^{2}/2}~H_{n}(y)~H_{m}(y)=\sqrt{2\pi} ~m!~e^{-x^{2}/2}~(ix)^{n-m}~L_{m}^{n-m}(x^{2}).
\label{eq:HL9}
\end{equation}
Similarly, if we put $\xi_{i}=0$ and write $\xi_{r}=x$ there, the formula will then become
\begin{equation}
\int_{-\infty}^{\infty} dy~e^{-y^{2}/2}~H_{n}(y+x)~H_{m}(y-x)=\sqrt{2\pi}~m!~(x)^{n-m}~L_{m}^{n-m}(x^{2}).
\label{eq:HL10}
\end{equation}
We note that (\ref{eq:HL10}) reduces, for the case $x=0$, to the well-known ortho-normality relation of Hermite polynomials. Furthermore, (\ref{eq:HL9}) also take the following forms: for $n-m=$even
\begin{equation}
\int_{0}^{\infty} dy~\cos xy~e^{-y^{2}/2}~H_{n}(y)H_{m}(y)=\sqrt{\frac{\pi}{2}} ~m!~e^{-x^{2}/2}~(ix)^{n-m}~L_{m}^{n-m}(x^{2}),
\end{equation}
and for $n-m=$odd
\begin{equation}
\int_{0}^{\infty} dy~\sin xy~e^{-y^{2}/2}~H_{n}(y)H_{m}(y)=-i~\sqrt{\frac{\pi}{2}} ~m!~e^{-x^{2}/2}~(ix)^{n-m}~L_{m}^{n-m}(x^{2}).
\end{equation}

Lastly we note that if in (\ref{eq:HL7}) $|x,y;+,->$'s are employed instead of $|x,y;+,+>$'s, then the resulting formula will be 
\begin{eqnarray}
&& \int_{-\infty}^{\infty} dx dy ~e^{i\epsilon xy/2}~e^{i\xi_{i} x} e^{-i\epsilon \xi_{r} y} e^{-(x^{2}+y^{2})/4}~H_{n}(x)H_{m}(y)  \nonumber  \\
&& \hspace{3cm}= 2\sqrt{2} \pi m! (i\epsilon)^{m} e^{-|\xi|^{2}/2} e^{i\xi_{i} \xi_{r}} \xi^{n-m} L_{m}^{n-m}(|\xi|^{2}).
\label{eq:Lagerre4}
\end{eqnarray}
When the $y-$integration is carried out in the above and the formula
\begin{equation}
\int_{-\infty}^{\infty} dy~e^{ix y}~e^{-y^{2}/4}~H_{m}(y)=2\sqrt{\pi}~e^{-x^{2}}~i^{m} H_{m}(2x)
\label{eq:Lagerre5}
\end{equation}
is used, (\ref{eq:Lagerre4}) will be reduced to (\ref{eq:HL8}). For the case $n=m$ and $\xi_{r}=\xi_{i}=0$, for example, (\ref{eq:Lagerre4}) will be 
\begin{equation}
\int _{-\infty}^{\infty} dx~dy~e^{ixy/2}~e^{-(x^{2}+y^{2})/4} ~H_{n}(x) H_{n}(y)=2\sqrt{2} \pi n! i^{n},
\end{equation}
which can also be obtained by use of (\ref{eq:Lagerre5}) together with the ortho-normality relation of Hermite polynomials.

\section{Eigenvalue problems of field operators: the wave-representation}
Having made the preliminary studies in the foregoing sections we now proceed to our main theme, i.e., the eigenvalue problems of $\phi(\vec{x},0)$ and $\pi(\vec{x},0)$ given respectively by (\ref{eq:phi2}) and (\ref{eq:pi2}). As already mentiond in sect 1, simultaneous eigenstates of $\phi(\vec{x},0)$'s or of $\pi(\vec{x},0)$'s at all spatial points $\vec{x}$ are given as direct products of the respective eigenstates of the operators $(a(0) \pm a^{\dagger}(0))$ and of $(a(\vec{k}) \pm a^{\dagger}(-\vec{k}))$'s with $\vec{k} \in K$, where upper (lower) signs are for $\phi(\vec{x},0) (\pi(\vec{x},0))$.

\subsection{Basis vectors}

 First let us  consider the case of $\phi(\vec{x},0)$. For the operator $(a(0)+a^{\dagger}(0))$ of the mode $\vec{k}=0$, the eigenstate $|\xi(0)>$ with eigenvalue $\xi(0)$ (real) is obtained from (\ref{eq:state1})   by putting $\epsilon=+$, $\xi=\xi(0)$, $a=a(0)$, and replacing $|0>$ with $|0>_{p}$:
\begin{equation}
|\xi(0)>\equiv \frac{1}{(2\pi)^{1/4}}e^{-|\xi(0)|^{2}/4}e^{-(a^{\dagger}(0))^{2}/2}e^{\xi(0)a^{\dagger}(0)}|0>_{p},
\label{eq:phi01}
\end{equation}
where $|0>_{p}$ is the state for which $a(\vec{k})|0>_{p}=0$ for all $\vec{k}$'s concerned and $_{p}<0|0>_{p}=1$. On the other hand, for the operators $(a(\vec{k}_{\ell})+a^{\dagger}(-\vec{k}_{\ell}))$ and $(a(-\vec{k}_{\ell})+a^{\dagger}(\vec{k}_{\ell}))$  of the mode $\vec{k}_{\ell} \in K$ their simultaneous eigenstate $|\xi(\vec{k}_{\ell})>$ having eigenvalues $\xi(\vec{k}_{\ell})$ and $\xi^{\ast}(\vec{k}_{\ell})$ respectively is obtained from (\ref{eq:state3})   by putting $\epsilon=+$, $\xi=\xi(\vec{k}_{\ell}),~a=a(\vec{k}_{\ell})$, $b=a(-\vec{k}_{\ell})$ and replacing $|0>$ with $|0>_{p}$:
\begin{equation}
|\xi(\vec{k}_{\ell})>\equiv \frac{1}{\sqrt{\pi}}e^{-|\xi(\vec{k_{\ell}})|^{2}/2} e^{- a^{\dagger}(\vec{k_{\ell}})a^{\dagger}(-\vec{k_{\ell}})}e^{\xi(\vec{k_{\ell}})a^{\dagger}(\vec{k_{\ell}})+ \xi^{\ast}(\vec{k_{\ell}})a^{\dagger}(-\vec{k_{\ell}})}~|0>_{p}.
\label{eq:phik1}
\end{equation}

Thus, if we form a direct product such that 
\begin{eqnarray}
&&|\xi(0);\xi(\vec{k_{1}}),\xi(\vec{k}_{2}),\cdots>\equiv |\xi(0)>\otimes |\xi(\vec{k_{1}})> \otimes |\xi(\vec{k_{2}})> \otimes \cdots  \nonumber  \\
&& \hspace{2cm}={\cal N} \exp\Bigl(-\frac{|\xi(0)|^{2}}{4}\Bigr)\times \exp\Bigl(-\frac{(a^{\dagger}(0))^{2}}{2}\Bigr)\times \exp\Bigl(\xi(0)a^{\dagger}(0)\Bigr)  \nonumber  \\
&&\hspace{2cm}\times \exp\Bigl(-\sum_{\vec{k_{\ell}}\in K}\frac{|\xi(\vec{k_{\ell}})|^{2}}{2}\Bigr) \times \exp\Bigl(-\sum_{\vec{k_{\ell}}\in K} a^{\dagger}(\vec{k_{\ell}})a^{\dagger}(-\vec{k_{\ell}}) \Bigr)  \nonumber  \\
&&\hspace{2cm}\times \exp\Bigl(\sum_{\vec{k_{\ell}}\in K}(\xi(\vec{k_{\ell}})a^{\dagger}(\vec{k_{\ell}})+ \xi^{\ast}(\vec{k_{\ell}})a^{\dagger}(-\vec{k_{\ell}})\Bigr)  |0>_{p}
\label{eq:phi0k1}
\end{eqnarray}
with ${\cal N}$ being a numerical factor, then the following eigenvalue equation will hold true:
\begin{equation}
\phi(\vec{x},0)~|\xi(0);\xi(\vec{k_{1}}),\xi(\vec{k}_{2}),\cdots>=\varphi_{\xi}(\vec{x})|\xi(0);\xi(\vec{k}_{1}),\xi(\vec{k}_{2}),\cdots>,
\label{eq:phieigeneq-1}
\end{equation}
where the eigenvalue $\varphi_{\xi}(\vec{x})=\varphi^{\ast}_{\xi}(\vec{x})$ is given as
\begin{eqnarray}
&&\varphi_{\xi}(\vec{x})=\sqrt{\frac{\hbar c^{2}}{2\omega_{0} L^{3}}}\xi(0)+\sum_{k_{\ell} \in K}\sqrt{\frac{\hbar c^{2}}{2\omega_{\ell}L^{3}}}\Bigl(\xi(k_{\ell})e^{i\vec{k}_{\ell}\vec{x}}+\xi^{\ast}(k_{\ell})e^{-i\vec{k}_{\ell}\vec{x}}\Bigr) \nonumber  \\
&&\hspace{1.5cm}=\sqrt{\frac{\hbar c^{2}}{2\omega_{0} L^{3}}}\xi(0)+2\sum_{k_{\ell} \in K}\sqrt{\frac{\hbar c^{2}}{2\omega_{\ell}L^{3}}}\Bigl(\xi_{r}(k_{\ell})\cos \vec{k}_{\ell}\vec{x}-\xi_{i}(k_{\ell})\sin \vec{k}_{\ell}\vec{x}\Bigr), 
\label{eq:ba1}
\end{eqnarray}
where $\omega_{\ell}=\omega_{k}|_{k=|\vec{k}_{\ell}|}$.
The function $\varphi_{\xi}(\vec{x})$ here may be interpreted as representing the matter wave associated with  the eigenstate concerned.

The case of $\pi(\vec{x},0)$ can be dealt with in a similar manner. This time we have only to make use of (\ref{eq:state1}) and (\ref{eq:state3}) with $\epsilon=-$, and the final result is as follows. The states defined by
\begin{eqnarray}
&&|\eta(0);\eta(\vec{k_{1}}),\eta(\vec{k}_{2}),\cdots>\equiv |\eta(0)>\otimes |\eta(\vec{k_{1}})> \otimes |\eta(\vec{k_{2}})> \otimes \cdots  \nonumber  \\
&& \hspace{2cm}={\cal N} ^{\prime}\exp\Bigl(-\frac{|\eta(0)|^{2}}{4}\Bigr)\times \exp\Bigl(\frac{(a^{\dagger}(0))^{2}}{2}\Bigr)\times \exp\Bigl(i\eta(0)a^{\dagger}(0)\Bigr) \nonumber  \\
&&\hspace{2cm} \times \exp\Bigl(-\sum_{\vec{k_{\ell}}\in K}\frac{|\eta(\vec{k_{\ell}})|^{2}}{2}\Bigr) \times \exp\Bigl(\sum_{\vec{k_{\ell}}\in K} a^{\dagger}(\vec{k_{\ell}})a^{\dagger}(-\vec{k_{\ell}}) \Bigr) \nonumber  \\
&&\hspace{2cm}\times \exp\Bigl(\sum_{\vec{k_{\ell}}\in K}(\eta(\vec{k_{\ell}})a^{\dagger}(\vec{k_{\ell}})- \eta^{\ast}(\vec{k_{\ell}})a^{\dagger}(-\vec{k_{\ell}})\Bigr)  |0>_{p} 
\label{eq:pistate1}
\end{eqnarray}
with ${\cal N^{\prime}}$ being a numerical factor will then satisfy the eigenvalue equation:
\begin{equation}
\pi(\vec{x},0)|\eta(0);\eta(\vec{k_{1}}),\eta(\vec{k}_{2}),\cdots>=\chi_{\eta}(\vec{x},0)|\eta(0);\eta(\vec{k}_{1}),\eta(\vec{k}_{2}),\cdots>,
\label{eq:pieigeneq-1}
\end{equation}
where the eigenvalue $\chi_{\eta}(\vec{x})= \chi^{\ast}_{\eta}(\vec{x})$ is given as
\begin{eqnarray}
&&\chi_{\eta}(\vec{x})=\sqrt{\frac{\hbar c^{2}}{2\omega_{0} L^{3}}}\eta(0)-i\sum_{k_{\ell}\in K} \sqrt{\frac{\hbar c^{2}}{2\omega_{\ell} L^{3}}}\Bigl(\eta(k_{\ell}) e^{i\vec{k_{\ell}}\vec{x}}-\eta^{\ast}(k_{\ell})e^{-i\vec{k}_{\ell}\vec{x}}\Bigr)  \nonumber  \\
&&\hspace{1.3cm}=\sqrt{\frac{\hbar c^{2}}{2\omega_{0} L^{3}}}\eta(0)+2\sum_{k_{\ell}\in K} \sqrt{\frac{\hbar c^{2}}{2\omega_{\ell} L^{3}}}\Bigl(\eta_{r}(k_{\ell}) \sin \vec{k_{\ell}}\vec{x}+\eta_{i}(k_{\ell})\cos \vec{k}_{\ell}\vec{x}\Bigr).
\label{eq:pieigenfunction}
\end{eqnarray}

The state vectors (\ref{eq:phi0k1}) have ortho-normality and completeness such that 
\begin{eqnarray}
&&<\xi^{\prime}(0);\xi^{\prime}(\vec{k_{1}}),\xi^{\prime}(\vec{k}_{2}),\cdots|\xi(0);\xi(\vec{k_{1}}),\xi(\vec{k}_{2}),\cdots> \nonumber  \\
&&\hspace{1.5cm}=\delta\bigl(\xi(0)-\xi^{\prime}(0)\bigr)~\delta\bigl(\xi(\vec{k}_{1})-\xi^{\prime}(\vec{k}_{1})\bigr) ~\delta\bigl(\xi(\vec{k}_{2})-\xi^{\prime}(\vec{k}_{2})\bigr)\cdots, \label{eq:baortho} \\
&& \int {\cal D}\xi ~|\xi(0);\xi(\vec{k_{1}}),\xi(\vec{k}_{2}),\cdots><\xi(0);\xi(\vec{k_{1}}),\xi(\vec{k}_{2}),\cdots|={\rm I}, \label{ba:complete}
\end{eqnarray}
where $\delta(\xi(\vec{k}))\equiv \delta(\xi_{r}(\vec{k}))\delta(\xi_{i}(\vec{k}))$ and ${\cal D}\xi \equiv d\xi(0)~ d^{2}\xi(\vec{k}_{1})~d^{2}\xi(\vec{k}_{2}) \cdots$.  This is because each factor state thereof satisfies (\ref{eq:ortho1}) and (\ref{eq:complete1}), as well as (\ref{eq:ortho3}) and (\ref{eq:complete3}). The same holds true, of course, of the state vectors (\ref{eq:pistate1}). The whole of (\ref{eq:phi0k1}) or of (\ref{eq:pistate1}) thus constitutes the basis vectors of what we have called in sect 1 the  wave-representaion of the quantized field concerned. The transformation function to connect this with the usual Fock or  particle-representation is given as the product of those of type (\ref{eq:H1}) and (\ref{eq:nm}).

An arbitrary state $|~>$ can be expressed in terms of, e.g., (\ref{eq:phi0k1}) as \begin{equation}
|~>=\int {\cal D}\xi ~\Psi(\xi(0); \xi(\vec{k}_{1}),\xi(\vec{k}_{2}),\cdots)~|\xi(0);\xi(\vec{k}_{1}),\xi(\vec{k}_{2}),\cdots>
\label{eq: astate}
\end{equation}
with $\Psi(\xi(0); \xi(\vec{k}_{1}),\xi(\vec{k}_{2}),\cdots)=<\xi(0); \xi(\vec{k}_{1}),\xi(\vec{k}_{2}),\cdots|~>$. Further, how the operators $\phi(\vec{x},0)$, $\pi(\vec{x},0)$ and their spatial derivatives act on $|>$ or on the probability amplitude $\Psi(\xi(0); \xi(\vec{k}_{1}),\xi(\vec{k}_{2}),\cdots)$ can be seen from the following relations:
\begin{eqnarray}
&& \phi(\vec{x},0)|~>=\int {\cal D}\xi~[ \varphi_{\xi}(\vec{x})~\Psi(\xi(0); \xi(\vec{k}_{1}),\xi(\vec{k}_{2}),\cdots)]|\xi(0); \xi(\vec{k}_{1}),\xi(\vec{k}_{2}),\cdots>,  \nonumber  \\
&&\pi(\vec{x},0) |~>= \int {\cal D}\xi ~\Bigl[\frac{1}{i}\frac{\delta}{\delta \varphi_{\xi}(\vec{x})}\Psi(\xi(0); \xi(\vec{k}_{1}),\xi(\vec{k}_{2}),\cdots)\Bigr]|\xi(0); \xi(\vec{k}_{1}),\xi(\vec{k}_{2}),\cdots>, \nonumber \\
\end{eqnarray}
where $\varphi_{\xi}(\vec{x})$ is given by (\ref{eq:ba1}) and 
\begin{eqnarray}
&&\frac{\delta}{\delta \varphi_{\xi}(\vec{x})} =2\sqrt{\frac{\hbar \omega_{0}}{2c^{2} L^{3}}}\frac{\partial}{\partial \xi(0)}+\sum_{\vec{k_{\ell}}\in K} \sqrt{\frac{\hbar \omega_{k_{\ell}}}{2c^{2} L^{3}}}\Bigl\{e^{i\vec{k_{\ell}}\vec{x}}\Bigl(\frac{\partial}{\partial \xi_{r}(k_{\ell})}+i\frac{\partial}{\partial \xi_{i}(k_{\ell})}\Bigr) \nonumber \\
&&\hspace{5.5cm}+e^{-i\vec{k_{\ell}}\vec{x}}\Bigl(\frac{\partial}{\partial \xi_{r}(k_{\ell})}-i\frac{\partial}{\partial \xi_{i}(k_{\ell})}\Bigr)\Bigr\}    \nonumber \\
&&\hspace{1.5cm}=2\Bigl[\sqrt{\frac{\hbar \omega_{0}}{2c^{2} L^{3}}}\frac{\partial}{\partial \xi(0)}+\sum_{\vec{k_{\ell}} \in K} \sqrt{\frac{\hbar \omega_{k_{\ell}}}{2c^{2} L^{3}}}\Bigl\{\cos\vec{k_{\ell}}\vec{x}~\frac{\partial}{\partial \xi_{r}(k_{\ell})} \nonumber \\
&&\hspace{5.5cm}- \sin\vec{k_{\ell}}\vec{x}~\frac{\partial}{\partial \xi_{i}(k_{\ell})}\Bigr\} \Big].
\end{eqnarray}
In the above we have taken account of the relations (\ref{eq:ab-1}) and (\ref{eq:ab}).

Incidentally, the expectation value of $\phi(\vec{x},0)$ for the state (\ref{eq: astate}) is written as 
\begin{equation}
<~|\phi(\vec{x},0)|~>=\Phi(0)+\sum_{\vec{k}\in K}~\Bigl(\Phi(\vec{k}) e^{i\vec{k}\vec{x}}+\Phi^{\ast}(\vec{k}) e^{-i\vec{k}\vec{x}}\Bigr),
\label{eq:hiex}
\end{equation}
where
\begin{eqnarray}
&& \Phi(0) \equiv \sqrt{\frac{\hbar c^{2}}{2\omega L^{3}}}~\int {\cal D}\xi~\xi(0) |\Psi(\xi(0); \xi(\vec{k}_{1}),\xi(\vec{k}_{2}),\cdots)|^{2},  \\
&&\Phi(\vec{k}) \equiv \sqrt{\frac{\hbar c^{2}}{2\omega L^{3}}}~\int {\cal D}\xi~\xi(\vec{k}) |\Psi(\xi(0); \xi(\vec{k}_{1}),\xi(\vec{k}_{2}),\cdots)|^{2}.
\label{eq:baexp}
\end{eqnarray}
The above quantity (\ref{eq:hiex}) may be interpreted as the matter wave associated with the state $|~>$.

\subsection{Quantum representatives of classical fields}
The basis vectors (\ref{eq:phi0k1}) or (\ref{eq:pistate1}) are not nomalized to unity, however. For evaluating physical quantities it is more convenient therefore to employ instead  Gaussian wave-packets with width $d$ $(>0)$ around the point $(\bar{\xi}(0);  \bar{\xi}(\vec{k}_{1}), \bar{\xi}(\vec{k}_{2}), \cdots)$ such that
\begin{eqnarray}
&&|\bar{\xi}(0); \bar{\xi}(\vec{k}_{1}),\bar{\xi}(\vec{k}_{2}),\cdots;d \gg \nonumber  \\
&&\hspace{1cm} = \int {\cal D}\xi ~\Psi_{\bar{\xi}}(\xi(0);\xi(k_{1}),\xi(k_{2}),\cdots)~|\xi(0);\xi(\vec{k}_{1}),\xi(\vec{k}_{2}),\cdots> \nonumber  \\
&& \hspace{1cm} \equiv |\bar{\xi};d \gg
\label{eq:Gbasis1}
\end{eqnarray}
with
\begin{eqnarray}
&&\Psi_{\bar{\xi}}(\xi(0);\xi(\vec{k}_{1}),\xi(\vec{k}_{2}),\cdots)  \nonumber  \\
&&\hspace{2cm}\equiv \Bigl(\frac{1}{2\pi d}\Bigr)^{1/4}~\exp\Bigl(-\frac{|\xi(0)-\bar{\xi}(0)|^{2}}{4d}\Bigr)  \nonumber  \\
&&\hspace{3cm}\times \prod_{\vec{k_{\ell}}\in K} \frac{1}{\sqrt{\pi d}}\exp\Bigl(-\frac{|\xi(\vec{k}_{\ell})-\bar{\xi}(\vec{k}_{\ell})|^{2}}{2d}\Bigr),  \label{eq:Gbasis2}
\end{eqnarray}
and similarly for the case of (\ref{eq:pistate1}). The scalar product of two such states is then found to be 
\begin{equation}
\ll \bar{\xi}^{\prime};d|\bar{\xi};d \gg =e^{-|\bar{\xi}(0)-\bar{\xi}^{\prime}(0)|^{2}/8d}~\prod_{\vec{k}_{\ell}\in K} e^{-|\bar{\xi}(\vec{k}_{\ell})-\bar{\xi}^{\prime}(\vec{k}_{\ell})|^{2}/4d},  \label{eq:innnerproduct1}
\end{equation}
showing that the two states are approximately orthogonal only when some of the $(|\bar{\xi}-\bar{\xi}^{\prime}|^{2}/d)$'s are far greater than unity.

Let us now evaluate the expectation value of $\phi(\vec{x},0)$ with respect to the state (\ref{eq:Gbasis1}). By using (\ref{eq:hiex}) $\sim$ (\ref{eq:Gbasis2}) together with the relation $\varphi_{\xi+\xi^{\prime}}(\vec{x})=\varphi_{\xi}(\vec{x})+\varphi_{\xi^{\prime}}(\vec{x})$ we find
\begin{eqnarray}
&& \ll \bar{\xi};d|\phi(\vec{x},0)|\bar{\xi};d \gg = \varphi_{\bar{\xi}}(\vec{x}) \nonumber  \\
&& \hspace{0.5cm}=\sqrt{\frac{\hbar c^{2}}{2\omega_{0} L^{3}}}\bar{\xi}(0)+2\sum_{k_{\ell}\in K} \sqrt{\frac{\hbar c^{2}}{2\omega_{\ell} L^{3}}}\Bigl(\bar{\xi}_{r}(k_{\ell}) \cos\vec{k_{\ell}}\vec{x}-\bar{\xi}_{i}(k_{\ell}) \sin\vec{k_{\ell}}\vec{x}\Bigr).  \nonumber  \\
\label{eq:barxi1}
\end{eqnarray}

Thus, given a classical field $\varphi_{cl}(\vec{x})=\varphi_{\bar{\xi}}(\vec{x})$ we Fourier-analyse it to determine its Fourier coefficients. From (\ref{eq:barxi1}) we can then determine the $\bar{\xi}(0)$ and $\bar{\xi}(\vec{k}_{\ell})$'s, and then from (\ref{eq:Gbasis1}) the corresponding state $|\bar{\xi};d\gg$. In this case, we then have $\ll \bar{\xi};d|\phi(\vec{x},0)|\bar{\xi};d \gg=\varphi_{cl}(\vec{x})$. More generally, if any state $|~>$ satisfies the relation $<~|\phi(\vec{x},0)|~>=\varphi_{cl}(\vec{x})$, we shall call this $|>$  a {\it quantum representative} of $\varphi_{cl}(\vec{x})$. Thus, the above  $|\bar{\xi};d\gg$  is one of such states.

The operator $\pi(\vec{x},0)$, on the other hand acts on $|\bar{\xi};d \gg$ as follows:
\begin{equation}
\pi(\vec{x},0)|\bar{\xi};d \gg=\int {\cal D}\xi ~\pi_{\xi-\bar{\xi}}(\vec{x})~\Psi_{\bar{\xi}}(\xi(0);\xi(\vec{k}_{1}),\xi(\vec{k}_{1}),\cdots)~|\xi(0);\xi(\vec{k}_{1}),\xi(\vec{k}_{2}),\cdots>,
\end{equation}
where
\begin{equation}
\pi_{\xi}(\vec{x})\equiv \frac{i}{d}\Bigl[\sqrt{\frac{\hbar \omega_{0}}{2c^{2}L^{3}}}\xi(0)+2\sum_{\vec{k}_{\ell} \in K} \sqrt{\frac{\hbar \omega_{\ell}}{2c^{2}L^{3}}} \Bigl\{\xi_{r}(\vec{k}_{\ell})\cos \vec{k}_{\ell}\vec{x}- \xi_{i}(\vec{k}_{\ell})\sin \vec{k}_{\ell}\vec{x}\Bigr\}\Bigr].
\end{equation}
It thus follows that
\begin{equation}
\ll \bar{\xi};d|\pi(\vec{x},0)|\bar{\xi};d \gg=0.
\end{equation}

Lastly we collect the expectation values of some relevant quantities in the limit $K \to \bar{K}$ and $L\to \infty$:
\begin{eqnarray}
&&\ll \bar{\xi};d|\phi(\vec{x}_{1},0)\phi(\vec{x}_{2},0)|\bar{\xi};d\gg=\varphi_{\bar{\xi}}(\vec{x}_{1})\varphi_{\bar{\xi}}(\vec{x}_{2})+ \frac{\hbar d}{4\pi^{2}}\frac{\mu}{|\vec{x}_{1}-\vec{x}_{2}|}K_{1}(\mu |\vec{x}_{1}-\vec{x}_{2}|),  \nonumber  \\
&&  \\
&&\ll \bar{\xi}:d|\pi(\vec{x}_{1},0)\pi(\vec{x}_{2},0)|\bar{\xi};d \gg=-\frac{\hbar}{4\pi^{2} c^{3}d}\Bigl(\frac{\mu}{|\vec{x}_{1}-\vec{x}_{2}|}\Bigr)^{2} ~K_{2}(\mu|\vec{x}_{1}-\vec{x}_{2}|),  \nonumber  \\
&&  \\
&&\ll \bar{\xi}:d|\pi(\vec{x}_{1},0)\phi(\vec{x}_{2},0)|\bar{\xi};d \gg =-\frac{i \hbar}{2}\delta(\vec{x}_{1}-\vec{x}_{2}),
\end{eqnarray}
where  $K_{\nu}$'s denote modified Bessel functions.

Further the expectation value of the total number operator reads
\begin{equation}
\ll \bar{\xi};d|\sum_{\vec{k}\in K} a^{\dagger}(\vec{k})a(\vec{k})|\bar{\xi};d\gg=\frac{1}{4}\sum_{\vec{k}\in K}|\bar{\xi}(\vec{k})|^{2}+\frac{1}{4}\sum_{\vec{k}\in K}\Bigl(\sqrt{d}-\frac{1}{\sqrt{d}}\Bigr)^{2}
\end{equation}
where we have put $\bar{\xi}(-\vec{k})\equiv \bar{\xi}^{\ast}(\vec{k})$ in accordance with (\ref{eq:ba1}) and (\ref{eq:barxi1}). The last result implies that the total number of particles is proportional, apart  from an additional  constant, to the absolute square of the corresponding classical field $\varphi_{cl}(\vec{x})=\varphi_{\bar{\xi}}(\vec{x})$ given by (\ref{eq:barxi1}).

\section{Two kinds of vacua}

It is usual in literature to define the vacuum state as the one containing no particles therein, or in our terminology, as the ground state in the particle-representation. As mentioned in sect 1 let us  call this state the particle-vacuum $|0>_{p}$. Similarly, the wave-representation is expected to have its own ground state, or the wave-vacuum $|0>_{w}$. This should be the one in which the field variables $\phi(\vec{x},0)$, loosely speaking, completely vanishes. More precisely, it is to be defined as the eigenstate of $\phi(\vec{x},0)$ with eigenvalue $\varphi_{\xi}(\vec{x})=0$, that is,
\begin{equation}
|0>_{w}\equiv |\xi(0);\xi(\vec{k}_{1}), \xi(\vec{k}_{2}), \cdots> \mid _{{\rm all}~\xi(\vec{k})'{\rm s}=0}.
\label{eq:vw}
\end{equation}

It is then easy to show that the basis vectors, e.g., (\ref{eq:phi0k1}), in the wave-representation can be created from $|0>_{w}$ as follows:
\begin{equation}
|\xi(0);\xi(\vec{k}_{1}),\xi(\vec{k}_{2}),\cdots>= D_{0}(\xi(0))~\prod_{\vec{k}_{\ell}\in K} D_{\vec{k}_{\ell}} (\xi(\vec{k}_{\ell}))|0>_{w},
\end{equation}
where $D_{0}$ and $D_{\vec{k}}$ are unitary operators such that
\begin{eqnarray}
&& D_{0}(\xi) \equiv \exp \Bigl[\frac{1}{2}\Bigl(\xi a^{\dagger}(0)-\xi^{\ast} a(0)\Bigr)\Bigr],  \\
&& D_{\vec{k}}(\xi) \equiv \exp \Bigl[\frac{1}{2}\Bigl(\xi a^{\dagger}(\vec{k})+\xi^{\ast} a^{\dagger}(-\vec{k})-\xi^{\ast} a(\vec{k})-\xi a(-\vec{k})\Bigr)\Bigr],
\end{eqnarray}
and satisfy 
\begin{equation}
D_{0}(\xi)D_{0}(\xi^{\prime})=D_{0}(\xi+\xi^{\prime}), ~~~~D_{\vec{k}}(\xi)D_{\vec{k}}(\xi^{\prime})=D_{\vec{k}}(\xi+\xi^{\prime}).
\label{eq:prel}
\end{equation}
We note that the above operators play the role similar to the so-called coherent operators with which to create coherent states from $|0>_{p}$ (cf. for example \cite{DFW}) , but unlike these they have the property (\ref{eq:prel}) or the commutativity $[D_{0}(\xi), D_{0}(\xi^{\prime})]=[D_{\vec{k}}(\xi), D_{\vec{k}}(\xi^{\prime})]=0$.

Let us now turn to studying further properties of two kinds of vacua. When taking account of (\ref{eq:0state1}) and (\ref{eq:vacuum2}), we can  write the particle-vacuum $|0>_{p}$ as
\begin{equation}
|0>_{p}=\Bigl(\frac{1}{2\pi}\Bigr)^{1/4}\int {\cal D}\xi~e^{-|\xi(0)|^{2}/4}~\prod_{\vec{k}_{\ell}\in K}~\frac{1}{\sqrt{\pi}}~e^{-|\xi(\vec{k}_{\ell})|^{2}/2}~|\xi(0);\xi(\vec{k}_{1}),\xi(\vec{k}_{2}),\cdots >.
\label{eq:vpstate}
\end{equation}
The right-hand side of (\ref{eq:vpstate}) provides actually the exact and explicit expression for what is commonly known as the {\it vaccum fluctuation} of the field. It is worth noticing here that $|0>_{p}$ of the above form, (\ref{eq:vpstate}), coincides with
\begin{equation}
|0>_{p}=|0;0,0,\cdots;d\gg \mid_{d =1}.
\label{eq:vp2}
\end{equation}

As for the wave-vacuum, on the other hand, it is more realistic physically to adopt, instead of the above $|0>_{w}$, the wave-packet state (\ref{eq:Gbasis1}) with all $\bar{\xi}$'s $=0$, that is,
\begin{eqnarray}
|0\gg_{w} &\equiv& |0;0,0,\cdots;d \gg \mid_{d \ll 1} \nonumber  \\
&=&\Bigl(\frac{1}{4\pi d}\Bigr)^{1/4}\int {\cal D}\xi~e^{-|\xi(0)|^{2}/4d}~\prod_{\vec{k}_{\ell}\in K}~\Bigl(\frac{1}{\sqrt{\pi d}}~e^{-|\xi(\vec{k}_{\ell})|^{2}/2d}\Bigr)\mid_{d \ll 1}~ \nonumber  \\
&&\hspace{5cm}\times |\xi(0);\xi(\vec{k}_{1}),\xi(\vec{k}_{2}),\cdots >,  \label{eq;vwstate}
\end{eqnarray}
which is to be contrasted with (\ref{eq:vp2}).

In this connection it may 
be of interest to know about the particle contents in the state (\ref{eq:Gbasis1}). To do this we first limit ourselves to the case of mode $\vec{k}=0$, and denote by $|n>$ and $|\bar{\xi}\gg$ the state vectors for this mode in the particle- and wave-representations, respectively. By use of (\ref{eq:state1-3}) we find after some calculations that
\begin{equation}
<n|\bar{\xi} \gg =\sqrt{\frac{2\sqrt{d}}{n!(d+1)}}\Bigl(\frac{1-d}{1+d}\Bigr)^{n/2} ~e^{-\bar{\xi}^{2}(0)/4(d+1)}~H_{n}(\frac{\bar{\xi}(0)}{\sqrt{1-d^{2}}})~~~{\rm for} ~~0<d \le 1.
\label{eq:nxi}
\end{equation}
Incidentally, for $\bar{\xi}(0)=0$ this is non-vanishing only for $n=$ even, whereas for $d=1$ this reduces to $(\bar{\xi}^{2}(0)/2)^{n} \exp(-\bar{\xi}^{2}(0)/8)/\sqrt{n!}$, which in the case of $\bar{\xi}(0)=0$ is non-vanishing only for $n=0$, as expected from (\ref{eq:vp2}).

For the case of mode $(\vec{k},-\vec{k})$ with $\vec{k} \in K$ we similarly denote by $|n,m>$ and $|\bar{\xi}\gg$  the relevant parts of the basis vectors for this mode. By virtue of (\ref{eq:nm}) we obtain,  e.g., for $n-m \ge 0$:
\begin{equation}
<n,m|\bar{\xi}\gg =\frac{1}{\pi \sqrt{d}}\sqrt{\frac{m!}{n!}}(-1)^{m} ~e^{-|\bar{\xi}(\vec{k})|^{2}/2d} \sum_{s=0}^{\infty} ~\frac{1}{s! d^{s}}~ I_{1}^{(n,m,s)}~ I_{2}^{(n,m,s)},
\label{eq:nmxi}
\end{equation}
where
\begin{eqnarray}
&&  I_{1}^{(n,m,s)}\equiv \int_{0}^{r} d r ~r^{n-m+s+1} ~e^{-(d+1)r^{2}/d}~L_{m}^{n-m}(r^{2}), \nonumber  \\
&& I_{2}^{(n,m,s)}\equiv \int_{0}^{2\pi} d\theta ~e^{i(n-m)\theta} (\bar{\xi}_{r}(\vec{k})\cos \theta+\bar{\xi}_{i}(\vec{k})\sin\theta)^{s}.
\end{eqnarray}
In the case of $\bar{\xi}_{r}(\vec{k})=\bar{\xi}_{i}(\vec{k})=0$ corresponding to $|0\gg_{w}$, we find that $I_{2}=2\pi \delta_{nm}\delta_{s0}$. Then $ I_{1}$ for this case is easily calculated, to give the final result
\begin{equation}
<n,m|\bar{\xi} \gg \mid_{\bar{\xi}(\vec{k})=0}~=~\frac{2\sqrt{d}}{d+1}~\Bigl(\frac{d-1}{d+1}\Bigr)^{n}~ \delta_{nm}.
\label{eq:barxi0}
\end{equation}
This  implies that for the state $|0\gg_{w}$  the partcles appear only in pairs with opposite momenta $\pm \hbar \vec{k}$, and its probabilities are the same for all $\vec{k}$'s. This property of $|0\gg_{w}$ is also shared by $|0>_{w}$, as may be seen from (\ref{eq:vacuum3}). All the results obtained above shows that the two kind of vacua are of the nature mutually complementary.
 
We are now in a position to see the reason why we have had to restict the region of $\vec{k}_{\ell}$'s to $K$. The above results (\ref{eq:nxi}) and (\ref{eq:barxi0}) show that if $K$ is extended to the entire region $\bar{K}$, the inner product $|<0|0\gg_{w}|$ will take the form of an infinite product such as $r_{0} \prod _{\ell}r_{\ell}$ with all $r$'s being smaller than unity, hence $|<0|0 \gg_{w}| \to 0$. This in turn implies that the particle- and wave-representations tend to be inequivalent with each other, a situation we just wanted to avoid. The same is also true of $|~_{p}<0|0>_{w}|$ as seen from (\ref{eq:0state1}) and (\ref{eq:vacuum3}).

We end this section by making a physical remark. Physical results to be compared with experiments, such as (\ref{eq:nxi}) $\sim$ (\ref{eq:barxi0}), depend of course on the parameter $d$. Needless to say, its magnitude should be so chosen as to be far smaller than the accuracy in measuring the $\bar{\xi}(\vec{k})$'s.

\section{Further remarks}
\subsection{Two kinds of superposition}

In field theories one and the same word {\it superposition} is used in regards to two different objects, that is, classical fields (or waves) and quantum states. The two kinds of superposition are quite distinct concepts, and as shown below they naturally give rise to  differences in physical results. Thus, to make a distinction between the two let us hereafter use the words c- and q-superposition for the former and the latter, respectively.

In sect 5 (cf.(\ref{eq:phieigeneq-1})) we have interpreted the eigenvalues $\varphi_{\xi}(\vec{x})$ of the field operator $\phi(\vec{x},0)$ as representing the matter waves associated with the field. As already mentiond, it is clear, from the definition of $\varphi_{\xi}(\vec{x})$, (\ref{eq:ba1}), that 
\begin{equation}
\varphi_{\xi}(\vec{x})+\varphi_{\xi^{\prime}}(\vec{x})=\varphi_{\xi+\xi^{\prime}}(\vec{x}),
\label{eq:varphi12}
\end{equation}
which is a c-superposition of two c-number quantities or matter waves $\varphi_{\xi}(\vec{x})$ and $\varphi_{\xi^{\prime}}(\vec{x})$, to form a third such quantity $\varphi_{\xi+\xi^{\prime}}(\vec{x})$. On the other hand, we may consider another, q-superposition of the corresponding two eigenstates of $\phi(\vec{x},0)$, $|\xi(0);\xi(\vec{k}_{1}),\xi(\vec{k}_{2}),\cdots >$ and $|\xi^{\prime}(0);\xi^{\prime}(\vec{k}_{1}),\xi^{\prime}(\vec{k}_{2}),\cdots >$, having the eigenvalues $\varphi_{\xi}(\vec{x})$ and $\varphi_{\xi^{\prime}}(\vec{x})$ respectively. In this case, however, the result of  superposition is not equal in general to $|\xi(0)+\xi^{\prime}(0);\xi(\vec{k}_{1})+\xi^{\prime}(\vec{k}_{1}),\xi(\vec{k}_{2})+\xi^{\prime}(\vec{k}_{2}),\cdots >$ , having the eigenvalue $\varphi_{\xi+\xi^{\prime}}(\vec{x})$ (as is clear from the definition (\ref{eq:phi0k1})). That is to say, we have in general
\begin{eqnarray}
&& |\xi(0);\xi(\vec{k}_{1},\xi(\vec{k}_{2}),\cdots >+|\xi^{\prime}(0);\xi^{\prime}(\vec{k}_{1},\xi^{\prime}(\vec{k}_{2}),\cdots >  \nonumber  \\
&&\hspace{2cm}\ne |\xi(0)+\xi^{\prime}(0);\xi(\vec{k}_{1})+\xi^{\prime}(\vec{k}_{1}),\xi(\vec{k}_{2})+\xi^{\prime}(\vec{k}_{2}),\cdots >.
\label{eq:superpose}
\end{eqnarray}
This clearly indicates that the two kinds of superposition should strictly be distinguished from each other.

The inequality of the type (\ref{eq:superpose}) holds also for quantum representatives $|\bar{\xi};d\gg$ of classical waves $\varphi_{\bar{\xi}}(\vec{x})$. To see this let us consider two states $|1>$ and $|2>$ defined  by
\begin{eqnarray}
&& |1>\equiv \Bigl\{ |\bar{\xi};d \gg +|\bar{\xi}^{\prime};d\gg|\Bigr\}/\sqrt{2} (1+D)^{1/2},  \nonumber  \\
&& |2>\equiv |(\bar{\xi}+\bar{\xi}^{\prime})/2;d \gg,
\end{eqnarray}
where $D \equiv \ll \bar{\xi}^{\prime};d|\bar{\xi};d\gg~ >0$, and the two sates $|1>$ and $|2>$ are so adjusted that for the case $\bar{\xi}=\bar{\xi}^{\prime}$ we have $|1>=|2>=|\bar{\xi};d\gg$.

It is  then easy to show that the inequality $|1>\ne |2>$ by actually calculating some expectation values with respect to each of these states.   We have, for example,
\begin{eqnarray}
&&<1|\phi(\vec{x},0)|1>=<2|\phi(\vec{x},0)|2>=\frac{1}{2}\{\varphi_{\bar{\xi}}(\vec{x})+\varphi_{\bar{\xi}^{\prime}}(\vec{x})\},  \\
&&<1|\phi(\vec{x}_{1},0)\phi(\vec{x}_{2},0)|1>=<2|\phi(\vec{x}_{1},0)\phi(\vec{x}_{2},0)|2>   \nonumber  \\
&&\hspace{2cm}+\frac{1}{4(1+D)}\{\varphi_{\bar{\xi}}(\vec{x}_{1})-\varphi_{\bar{\xi}^{\prime}}(\vec{x}_{1})\}\{\varphi_{\bar{\xi}}(\vec{x}_{2})-\varphi_{\bar{\xi}^{\prime}}(\vec{x}_{2})\};  \\
&&<1|\pi(\vec{x},0)|1>=<2|\pi(\vec{x},0)|2>=0,  \\
&& <1|\pi(\vec{x}_{1},0)\pi(\vec{x}_{2},0)|1>=<2|\pi(\vec{x}_{1},0)\pi(\vec{x}_{2},0)|2>   \nonumber  \\
&&\hspace{2cm}-\frac{D}{4(1+D)}~{\cal \pi}_{(\bar{\xi}_{1}-\bar{\xi}^{\prime})}(\vec{x}_{1}){\cal \pi}_{(\bar{\xi}_{1}-\bar{\xi}^{\prime})}(\vec{x}_{2}) ;
\end{eqnarray}
thus for the energy density ${\cal H}(\vec{x})=\{c^{2}\pi^{2}(\vec{x},0)+(\vec{\nabla}\phi(\vec{x},0))^{2}+\mu^{2}\phi^{2}(\vec{x},0) \}/2$
\begin{eqnarray}
&&<1|{\cal H}(\vec{x})|1>=<2|{\cal H}(\vec{x})|2>  \nonumber  \\
&& \hspace{1cm}+\frac{1}{8(1+D)}\Bigl\{\Bigl(\vec{\nabla}\varphi_{(\bar{\xi}-\bar{\xi}^{\prime})}(\vec{x})\Bigr)^{2}+\mu^{2}\Bigl(\varphi_{(\bar{\xi}-\bar{\xi}^{\prime})}(\vec{x})\Bigr)^{2}-c^{2}D\Bigl({\cal \pi}_{(\bar{\xi}-\bar{\xi}^{\prime})}(\vec{x})\Bigr)^{2}\Bigr\} . \nonumber  \\
&&
\end{eqnarray}

The above is enough to show that $|1> \ne |2>$ in general,  hence the inequality of the type (\ref{eq:superpose}). It is also clear that the two states, $|1>$ and $|2>$, leads to a difference in the interference effects that arise originally from the c-superposition of $\varphi_{\bar{\xi}}(\vec{x})$ and $\varphi_{\bar{\xi}^{\prime}}(\vec{x})$. 

Summarizing, the lesson we learn from the above is that the c- and q- superposition should be kept distinct in discussions of quantum fields.

\subsection{Applications to the electromagnetic and other fields}

Let us first consider the case of electromagnetic fields. In the Coulomb gauge the vector potential $\vec{A}(\vec{x},0)$ is written as
\begin{equation}
\vec{A}(\vec{x},0)=\sqrt{\frac{2\pi \hbar c^{2}}{L^{3}}} \sum_{\vec{k}\in K} \sum_{j=1,2} ~\frac{\vec{e}(\vec{k},j)}{\sqrt{\omega_{k}}} \Bigl\{a(\vec{k},j) e^{i\vec{k}\vec{x}}+a^{\dagger}(\vec{k},j)e^{-i\vec{k}\vec{x}}\Bigr\}.
\label{eq:Afield}
\end{equation}
Here $\omega_{k}\equiv c|\vec{k}|$, and $\vec{e}(\vec{k},j)$'s are unit vectors perpendicular to each other as well as to $\vec{k}$. For $\vec{k}=0$ the vectors $\vec{e}(0,j)$'s are not definable, but the corresponding terms in $\vec{A}$, if any, may be eliminated by means of a gauge transformation whose  gauge function is of the form  $\overrightarrow{{\rm const}.}\cdot \vec{x}$. And for $\vec{k}\in K$ we may choose them in such a way that $\vec{e}(-\vec{k},j)=\vec{e}(\vec{k},j)$. Then $\vec{A}(\vec{x},0)$ can be cast in a form similar to (\ref{eq:phi2}):
\begin{eqnarray}
&&\vec{A}(\vec{x},0)=\sqrt{\frac{2\pi \hbar c^{2}}{L^{3}}} \sum_{\vec{k}\in K} \sum_{j=1,2} ~\frac{\vec{e}(\vec{k},j)}{\sqrt{\omega_{k}}} \Bigl\{\Bigl(a(\vec{k},j)+a^{\dagger}(-\vec{k},j)\Bigr) e^{i\vec{k}\vec{x}}   \nonumber  \\
&&\hspace{5cm}+\Bigl(a(-\vec{k},j)+a^{\dagger}(\vec{k},j)\Bigr)e^{-i\vec{k}\vec{x}}\Bigr\},
\end{eqnarray}
so that our method of the wave-representation developed for the operator $\phi(\vec{x},0)$  can be straightly applied to the operator $\vec{A}(\vec{x},0)$ in the above form. In this case  the corresponding expressions for the field strengths $\vec{E}(\vec{x},0)$ and $\vec{H}(\vec{x},0)$ become
\begin{eqnarray}
&&\vec{E}(\vec{x},0)=i\sqrt{\frac{2\pi \hbar }{L^{3}}}\sum _{\vec{k}\in K}\sum_{j=1,2} \sqrt{\omega_{k}}~\vec{e}(\vec{k},j)\Bigl\{\Bigl(a(\vec{k},j)-a^{\dagger}(-\vec{k},j)\Bigr)e^{i\vec{k}\vec{x}}  \nonumber  \\
&&\hspace{6cm}-\Bigl(a^{\dagger}(\vec{k},j)-a(-\vec{k},j)\Bigr)e^{-i\vec{k}\vec{x}}\Bigr\},  \\
&&\vec{H}(\vec{x},0)=i\sqrt{\frac{2\pi \hbar c^{2}}{L^{3}}}\sum _{\vec{k}\in K}\sum_{j=1,2} \frac{\{\vec{k}\times\vec{e}(\vec{k},j)\}}{\sqrt{\omega_{k}}}\Bigl\{\Bigl(a(\vec{k},j)+a^{\dagger}(-\vec{k},j)\Bigr)e^{i\vec{k}\vec{x}}  \nonumber  \\
&&\hspace{6cm}-\Bigl(a^{\dagger}(\vec{k},j)+a(-\vec{k},j)\Bigr)e^{-i\vec{k}\vec{x}}\Bigr\}.
\end{eqnarray}

Now the eigenstates $|\xi(\vec{k}_{1},j_{1}),\xi(\vec{k}_{2},j_{}),\cdots>$ of the operator $\vec{A}(\vec{x},0)$ can be obtained simply from (\ref{eq:phi0k1}) by making some obvious modifications. An arbitrary state vector $|~>$ is then written as
\begin{equation}
|~>=\int {\cal D}\xi \Psi(\xi(\vec{k}_{1},j_{1}),\xi(\vec{k}_{2},j_{2}),\cdots )~|\xi(\vec{k}_{1},j_{1}),\xi(\vec{k}_{2},j_{2}),\cdots >,
\end{equation}
where  $\int {\cal D}\xi$ means both the integration and summation over all $\vec{k}_{\ell}$'s and $j_{\ell}$'s. In this case various operators act on this state in the following manner.
\begin{equation}
\vec{A}(\vec{x},0) |>=\int {\cal D}\xi~ \vec{{\cal A}}_{\xi}(\vec{x})~\Psi(\xi(\vec{k}_{1},j_{1}), \xi(\vec{k}_{2},j_{2}), \cdots )|\xi(\vec{k}_{1},j_{1}),\xi(\vec{k}_{2},j_{2}), \cdots >,
\end{equation}
where
\begin{equation}
\vec{{\cal A}}_{\xi}(\vec{x})\equiv 2\sqrt{\frac{2\pi \hbar c^{2}}{L^{3}}}\sum _{\vec{k}\in K}\sum_{j=1,2} \frac{\vec{e}(\vec{k},j)}{\sqrt{\omega_{k}}}\Bigl\{\xi_{r}(\vec{k},j) \cos \vec{k}\vec{x}-\xi_{i}(\vec{k},j) \sin \vec{k}\vec{x}\Bigr\};
\end{equation}
\begin{eqnarray}
&&\vec{E}(\vec{x},0)|> =-2 \sqrt{\frac{2\pi \hbar }{L^{3}}}\sum _{\vec{k}\in K}\sum_{j=1,2} \sqrt{\omega_{k}}~\vec{e}(\vec{k},j)  \nonumber  \\
&& \hspace{3cm}\times \int {\cal D}\xi~ \Bigl[\Bigl\{\cos \vec{k}\vec{x} \frac{\partial}{\partial \xi_{i}(\vec{k},j)}+\sin \vec{k}\vec{x} \frac{\partial}{\partial \xi_{r}(\vec{k},j)}\Bigr\}   \nonumber  \\
&&\hspace{3.5cm}\times \Psi(\xi(\vec{k}_{1},j_{1}), \xi(\vec{k}_{2},j_{2}), \cdots )\Bigr]|\xi(\vec{k}_{1},j_{1}),\xi(\vec{k}_{2},j_{2}), \cdots >, \nonumber \\&&  \\
&&H(\vec{x},0)|>=\int {\cal D}\xi~ \vec{{\cal H}}_{\xi}(\vec{x})~\Psi(\xi(\vec{k}_{1},j_{1}), \xi(\vec{k}_{2},j_{2}), \cdots )  \nonumber  \\
&&\hspace{4cm}\times |\xi(\vec{k}_{1},j_{1}),\xi(\vec{k}_{2},j_{2}), \cdots >
\end{eqnarray}
with
\begin{eqnarray}
&&\vec{{\cal H}}_{\xi}(\vec{x})\equiv -2\sqrt{\frac{2\pi \hbar c^{2}}{L^{3}}}\sum _{\vec{k}\in K}\sum_{j=1,2} \frac{\{\vec{k}\times\vec{e}(\vec{k},j)\}}{\sqrt{\omega_{k}}}\Bigl\{\sin \vec{k}\vec{x}~\xi_{r}(\vec{k},j)+\cos\vec{k}\vec{x}~\xi_{i}(\vec{k},j)\Bigr\}  \nonumber  \\
&&\hspace{3.5cm}=\vec{\nabla}_{x}\times \vec{{\cal A}}_{\bar{\xi}}(\vec{x}).
\end{eqnarray}
 
Further, in a way similar to (\ref{eq:Gbasis1}) and (\ref{eq:Gbasis2}) we can also construct wave-packet states $|\bar{\xi};d\gg$:
\begin{equation}
|\bar{\xi};d\gg  \equiv \int {\cal D}\xi~ \Psi_{\bar{\xi}}(\xi(\vec{k}_{1},j_{1}), \xi(\vec{k}_{2},j_{2}), \cdots )|\xi(\vec{k}_{1},j_{1}),\xi(\vec{k}_{2},j_{2}), \cdots >
\end{equation}
with 
\begin{equation}
\Psi_{\bar{\xi}}(\xi(\vec{k}_{1},j_{1}), \xi(\vec{k}_{2},j_{2}), \cdots )\equiv \prod_{\vec{k}\in K}\prod_{j=1,2}\frac{1}{\sqrt{\pi d}}\exp \Bigl(-\frac{|\xi(\vec{k};j)-\bar{\xi}(\vec{k},j)|^{2}}{2d}\Bigr).
\end{equation}

In this connection, the following expectation values may be of interest:
\begin{eqnarray}
&&\ll \bar{\xi};d|\vec{A}(\vec{x},0)|\bar{\xi};d\gg= \vec{{\cal A}}_{\bar{\xi}} (\vec{x}),  \\
&&\ll \bar{\xi};d|\vec{E}(\vec{x},0)|\bar{\xi};d\gg=0,  \\
&&\ll \bar{\xi};d|\vec{H}(\vec{x},0)|\bar{\xi};d\gg=\vec{{\cal H}}_{\bar{\xi}}(\vec{x});  \\
&&\ll \bar{\xi};d|\bigl(\vec{E}(\vec{x}_{1},0)\cdot \vec{E}(\vec{x}_{2},0)\bigr)|\bar{\xi};d\gg =\frac{8\pi \hbar}{L^{3}}\frac{1}{d}\sum_{\vec{k}\in K}\omega_{k}\cos\vec{k}(\vec{x}_{1}-\vec{x}_{2}),  \label{eq:EE} \\
&&\ll \bar{\xi};d|\bigl(\vec{H}(\vec{x}_{1},0)\cdot \vec{H}(\vec{x}_{2},0)\bigr)|\bar{\xi};d\gg =\frac{8\pi \hbar}{L^{3}}d \sum_{\vec{k}\in K}\omega_{k}\cos\vec{k}(\vec{x}_{1}-\vec{x}_{2}) \nonumber  \\
&&\hspace{6.5cm}+\vec{{\cal H}}_{\bar{\xi}}(\vec{x}_{1})\cdot \vec{{\cal H}}_{\bar{\xi}}(\vec{x}_{2}) ;  
\end{eqnarray}
and for the energy density ${\cal E}(\vec{x})$
\begin{equation}
{\cal E}(\vec{x})=\frac{\hbar}{L^{3}}\sum_{\vec{k}\in K}\Bigl(d+\frac{1}{d}\Bigr)+\frac{1}{8\pi}\vec{{\cal H}}_{\bar{\xi}}(\vec{x})\cdot \vec{{\cal H}}_{\bar{\xi}}(\vec{x}).
\label{eq:energy}
\end{equation}
Needless to say, the terms involving $\sum_{\vec{k}}$ in (\ref{eq:EE})$\sim$ (\ref{eq:energy}) correspond to the respective wave-vacuum values in the limit $d \to 0$. Incidentally, the roles of $\vec{E}$ and $\vec{H}$ in the above relations are , so to speak, interchanged, if the basis vectors of the form (\ref{eq:pistate1}) are employed.

The two kinds of vacua and other properties can be argued in the same way as in the preceding sections. It is thus expected that our wave-representation will be useful, e.g., in quantum optics, especially in clarifying the wave aspect thereof.

Lastly we spend only a few words about the case of Fermi fields. It is, of course, possible to formally construct the wave-representation by use of Grassmann numbers \cite{YO,OSK}. Since, however, such numbers cannot directly be related to observable quantities, we should say that the wave-representation would not be so useful as in the case of Bose fields.

\begin{acknowledgments}
We wish to thank Drs. S. Sakoda and Y. Ohunki for valuable discussions.
\end{acknowledgments}

\appendix

\section{}

\subsection{Proofs of (\ref{eq:ortho1}) $\sim$ (\ref{eq:complete1})}

A repeated use of Hausdorff's formula allows us to make an anzatz: 
\begin{equation}
e^{\pm \frac{\lambda}{2}a^{2}}e^{-\frac{\lambda}{2}(a^{\dagger})^{2}}=e^{-\alpha_{\pm}(\lambda)(a^{\dagger})^{2}}e^{-\beta_{\pm}(\lambda)(a^{\dagger}a+aa^{\dagger})}e^{-\gamma_{\pm}(\lambda)a^{2}}.
\label{eq:sakoda1},
\end{equation}
which is justified, however, by showing that the coefficients $\alpha_{\pm}, \beta_{\pm}$ and $\gamma_{\pm}$ are uniquely determined. Here the parameter $\lambda$, serving below as a regulator, is assumed to be $-1<\lambda<1$. To show this we first differentiate (\ref{eq:sakoda1}) with respect to $\lambda$ and compare the coefficients of the operators $a^{2},~(aa^{\dagger}+a^{\dagger}a)$ and $(a^{\dagger})^{2}$ on both sides, thereby  obtaining the following equations:
\begin{eqnarray}
&& \dot{\gamma}_{\pm}~(1 \pm \lambda^{2})^{2}=\mp\frac{1\mp \lambda^{2}}{2}, \nonumber  \\
&&\dot{\beta}_{\pm}+\dot{\gamma}_{\pm} ~\lambda (1\pm \lambda^{2})=\pm \frac{\lambda}{2},  \nonumber  \\
&&\dot{\alpha}_{\pm}+4\dot{\beta}_{\pm}~\alpha_{\pm}+\dot{\gamma}_{\pm}\lambda^{2}=\frac{1}{2}.
\label{eq:sakoda2}
\end{eqnarray}
The solutions of these differential equations under the initial conditions $a_{\pm}(0)=0, \beta_{\pm}(0)=0$ and $\gamma_{\pm}(0)=0$ are
\begin{eqnarray}
\alpha_{\pm}(\lambda)= \mp \gamma_{\pm}(\lambda)=\frac{\lambda}{2(1 \pm \lambda^{2})}, ~~\beta_{\pm}(\lambda)=\frac{1}{2} \log (1 \pm \lambda^{2}). 
\label{eq:sakoda3}
\end{eqnarray}
The above method of deriving (\ref{eq:sakoda1}) is originally due to Sakoda \cite{S}. Using (\ref{eq:sakoda1}) we then have
\begin{eqnarray}
&&<0|e^{\xi^{\prime \ast}a}e^{\pm\frac{\lambda}{2}a^{2}}e^{-\frac{\lambda}{2}(a^{\dagger})^{2}}e^{\xi a^{\dagger}}|0> \nonumber  \\
&& \hspace{0.5cm} =<0|e^{\xi^{\prime}a}e^{-\alpha_{\pm}(\lambda)(a^{\dagger})^{2}}e^{-\beta_{\pm}(\lambda)(a^{\dagger}a+aa^{\dagger})}e^{-\gamma_{\pm}(\lambda)a^{2}}e^{\xi a^{\dagger}}|0>   \nonumber  \\
&&\hspace{0.5cm}=<0|e^{\xi^{\prime \ast}(a-2\alpha_{\pm}a^{\dagger})}e^{-\beta_{\pm}(a^{\dagger}a+aa^{\dagger})}e^{\xi(a^{\dagger}-2\gamma_{\pm} a)}|0>  \nonumber  \\
&&\hspace{0.5cm}=<0|e^{\xi^{\prime \ast}(a-2\alpha_{\pm}a^{\dagger})}e^{\xi(a^{\dagger}e^{-2\beta_{\pm}}-2\gamma_{\pm} a e^{2\beta_{\pm}})}|0> e^{-\beta_{\pm}}  \nonumber  \\
&&\hspace{0.5cm}=e^{-\alpha_{\pm}(\xi^{\prime \ast})^{2}-\gamma_{\pm}\xi^{2}}e^{\xi^{\prime \ast}\xi e^{-2\beta_{\pm}}}e^{-\beta_{\pm}}.
\label{eq:sakoda4}
\end{eqnarray}

Substituting into (\ref{eq:sakoda4}) the above $\alpha_{+},\beta_{+}$ and $\gamma_{+}$ and doing the same for $\alpha_{-},\beta_{-}$ and $\gamma_{-}$ we obtain  respectively 
\begin{equation}
<0|e^{\xi^{\prime \ast}a}e^{\frac{\lambda}{2}a^{2}}e^{-\frac{\lambda}{2}(a^{\dagger})^{2}}e^{\xi a^{\dagger}}|0>=\frac{1}{\sqrt{1+\lambda^{2}}}~\exp \Bigl\{\frac{-\lambda(\xi^{\prime \ast})^{2}+\lambda \xi^{2}}{2(1+\lambda^{2})}\Bigr\}\exp \Bigl(\frac{\xi^{\prime \ast}\xi}{1+\lambda^{2}}\Bigr),
\label{eq:sakoda5}
\end{equation}
and
\begin{equation}
<0|e^{\xi^{\prime \ast}a}e^{\frac{-\lambda}{2}a^{2}}e^{-\frac{\lambda}{2}(a^{\dagger})^{2}}e^{\xi a^{\dagger}}|0>=\frac{1}{\sqrt{1-\lambda^{2}}}~\exp \Bigl\{-\frac{\lambda(\xi^{\prime \ast})^{2}+\lambda \xi^{2}}{2(1-\lambda^{2})}\Bigr\}\exp \Bigl(\frac{\xi^{\prime \ast}\xi}{1-\lambda^{2}}\Bigr).
\label{eq:sakoda6}
\end{equation}

We can now show that (\ref{eq:sakoda6}) in the limit $\lambda \to \pm 1$ leads to $<\xi^{\prime},\epsilon|\xi;\epsilon>$ (apart from normalization). First for the case $\epsilon=+$, we put there $\lambda=1-\kappa, \xi=\xi_{r}$ and $\xi^{\prime}=\xi^{\prime}_{r}$ to find in the limit $\kappa \to 0$
\begin{eqnarray}
<0|e^{\xi^{\prime \ast}a}e^{-\frac{1}{2}a^{2}}e^{-\frac{1}{2}(a^{\dagger})^{2}}e^{\xi a^{\dagger}}|0>&=&\lim_{\kappa \to 0} \frac{1}{\sqrt{2\kappa}}~e^{-\frac{(\xi_{r}-\xi^{\prime}_{r})^{2}}{4\kappa}}~e^{\frac{(\xi_{r}^{2}+\xi^{\prime 2}_{r})}{4}} \nonumber  \\
&=&\sqrt{2\pi}e^{\frac{(\xi_{r}^{2}+\xi^{\prime 2}_{r})}{4}}\delta(\xi_{r}-\xi^{\prime}_{r}).
\label{eq:sakoda7}
\end{eqnarray}
Next, for the case $\epsilon=-$, we put $\lambda=-1+\kappa, \xi=i\xi_{i}$ and $\xi^{\prime}=i\xi^{\prime}_{i}$ to find similarly
\begin{eqnarray}
<0|e^{\xi^{\prime \ast}a}e^{\frac{1}{2}a^{2}}e^{\frac{1}{2}(a^{\dagger})^{2}}e^{\xi a^{\dagger}}|0>&=&\lim_{\kappa \to 0} \frac{1}{\sqrt{2\kappa}}~e^{-\frac{(\xi_{i}-\xi^{\prime}_{i})^{2}}{4\kappa}}~e^{\frac{(\xi_{i}^{2}+|\xi^{\prime 2}_{i})}{4}} \nonumber  \\
&=&\sqrt{2\pi}e^{\frac{(\xi_{i}^{2}+\xi^{\prime 2}_{i})}{4}}\delta(\xi_{i}-\xi^{\prime}_{i}) .
\label{eq:sakoda8}
\end{eqnarray}
Relations (\ref{eq:sakoda7}) and (\ref{eq:sakoda8}) lead to (\ref{eq:ortho1}) in the text.
Lastly, putting in (\ref{eq:sakoda5}) $\lambda \approx 1, \xi=\xi_{r}$ and $\xi^{\prime}=i\xi^{\prime}_{i}$,  we then find
\begin{equation}
<0|e^{\xi^{\prime \ast}a}e^{\frac{1}{2}a^{2}}e^{\frac{1}{2}(a^{\dagger})^{2}}e^{\xi a^{\dagger}}|0>=\frac{1}{\sqrt{2}}~e^{(\xi_{r}^{2}+\xi^{\prime 2}_{i})/4}~e^{-i\xi_{r}\xi^{\prime}_{i}/2},
\label{eq:sakoda9}
\end{equation}
thereby proving (\ref{eq:trans1}).

On the other hand, for proving (\ref{eq:complete1}) we proceed as follows:
\begin{eqnarray}
&&\int_{-\infty}^{\infty} d\xi_{r}|\xi_{r};+><\xi_{r};+| \nonumber \\
&&\hspace{0.5cm}=\frac{1}{\sqrt{2\pi}}\sum_{n,m=0}^{\infty} \frac{1}{\sqrt{n!m!}}|n><m|~\int_{-\infty}^{\infty} d\xi_{r} ~e^{-\xi_{r}^{2}/2}~H_{n}(\xi_{r}) H_{m}(\xi_{r})  \nonumber  \\
&&\hspace{0.5cm}= \frac{1}{\sqrt{2\pi}}\sum_{n,m=0}^{\infty} n! \sqrt{2\pi} \delta_{nm}|n><m|=\sum_{n=0}^{\infty} |n><n|={\rm I},
\label{eq:sakoda10}
\end{eqnarray}
where use is made of (\ref{eq:nstate2}). The case $\epsilon=-$ can be dealt with in the same way.

\subsection{Proofs of (\ref{eq:H1}) and (\ref{eq:H2})}
By using the expression 
\begin{eqnarray}
&&e^{-\frac{\epsilon}{2}(a^{\dagger})^{2}+\xi a^{\dagger}} \nonumber \\
&&\hspace{0.5cm}=\sum _{n=0}^{\infty} \frac{1}{n!}\Bigl(\xi a^{\dagger}-\frac{\epsilon}{2}(a^{\dagger})^{2}\Bigr)^{n} =\sum_{n=0}^{\infty} \frac{1}{n!}\sum_{r=0}^{n} ~_{n}C_{r} (\xi a^{\dagger})^{n-r}\Bigl(-\frac{\epsilon}{2}(a^{\dagger})^{2}\Bigr)^{r} \nonumber \\
&&\hspace{0.5cm}=\sum_{n=0}^{\infty} \frac{\xi^{n}}{n!} \sum_{r=0}^{n}~_{n}C_{r} \xi^{n-r}\Bigl(-\frac{\epsilon}{2}\Bigr)^{r}(a^{\dagger})^{n+r}=\sum_{n=0}^{\infty} \frac{\xi^{n}}{n!}\sum_{r=0}^{n} ~_{n}C_{r} \Bigl(-\frac{\epsilon}{2\xi}\Bigr)^{r} (a^{\dagger})^{n+r}, \nonumber \\
\label{eq:sakoda11}
\end{eqnarray}
we find
\begin{eqnarray}
<n|\xi;\epsilon>&=& \frac{e^{-|\xi|^{2}/4}}{(2\pi)^{1/4}}\frac{1}{\sqrt{n!}}<0|a^{n} e^{-\frac{\epsilon}{2}(a^{\dagger})^{2}+\xi a^{\dagger}}|0>  \nonumber  \\
&=&\frac{e^{-|\xi|^{2}/4}}{(2\pi)^{1/4}}\frac{1}{\sqrt{n!}}\sum_{m=0}^{\infty} \frac{\xi^{m}}{m!}\sum_{r=0}^{m} ~_{m}C_{r} \Bigl(-\frac{\epsilon}{2\xi}\Bigr)^{r} <0|a^{n}(a^{\dagger})^{m+r}|0>  \nonumber  \\
&=& \frac{e^{-|\xi|^{2}/4}}{(2\pi)^{1/4}}\frac{1}{\sqrt{n!}}\sum_{m=0}^{\infty} \frac{\xi^{m}}{m!}\sum_{r=0}^{m} ~_{m}C_{r} \Bigl(-\frac{\epsilon}{2\xi}\Bigr)^{r} n! ~\delta_{n,m+r}.
\label{eq:sakoda12}
\end{eqnarray}

Thus, for the cases $n={\rm even}=2k$ and $n={\rm odd}=2k+1$, (\ref{eq:sakoda12}) can be rewitten respectively as
\begin{eqnarray}
<2k|\xi;\epsilon>&=&\frac{e^{-|\xi|^{2}/4}}{(2\pi)^{1/4}}\sqrt{(2k)!}\Bigl\{ _{2k}C_{0}~\frac{\xi^{2k}}{(2k)!}+ _{2k-1}C_{1}~\frac{\xi^{2k-1}}{(2k-1)!}\Bigl(-\frac{\epsilon}{2\xi}\Bigr)+  \nonumber  \\
&&~~~~~~+~_{2k-2}C_{2}~\frac{\xi^{2k-2}}{(2k-2)!}\Bigl(-\frac{\epsilon}{2\xi}\Bigr)^{2}+\cdots +~_{2k-k}C_{k}~\frac{\xi^{2k-k}}{(2k-k)!}\Bigl(-\frac{\epsilon}{2\xi}\Bigr)^{k}  \nonumber  \\
&=&\frac{e^{-|\xi|^{2}/4}}{(2\pi)^{1/4}}\sqrt{(2k)!}\sum_{r=0}^{k} ~_{2k-r}C_{r}~\frac{\xi^{2k-r}}{(2k-r)!}\Bigl(-\frac{\epsilon}{2\xi}\Bigr)^{r}  \nonumber  \\
&=&\frac{e^{-|\xi|^{2}/4}}{(2\pi)^{1/4}}\sqrt{(2k)!}~\xi^{2k} \sum_{r=0}^{k} ~\frac{1}{ r! (2k-2r)!} \Bigl(-\frac{\epsilon}{2\xi^{2}}\Bigr)^{r},
\label{eq:sakoda12}
\end{eqnarray}
and
\begin{eqnarray}
&&<2k+1|\xi;\epsilon> \nonumber \\
&&\hspace{0.5cm}=\frac{e^{-|\xi|^{2}/4}}{(2\pi)^{1/4}}\sqrt{(2k+1)!}\Bigl\{ _{2k+1}C_{0}~\frac{\xi^{2k+1}}{(2k+1)!}+ _{2k}C_{1}~\frac{\xi^{2k}}{(2k)!}\Bigl(-\frac{\epsilon}{2\xi}\Bigr)+  \nonumber  \\
&&\hspace{1cm}+~_{2k-1}C_{2}~\frac{\xi^{2k-1}}{(2k-1)!}\Bigl(-\frac{\epsilon}{2\xi}\Bigr)^{2}+\cdots +~_{2k+1-k}C_{k}~\frac{\xi^{2k+1-k}}{(2k+1-k)!}\Bigl(-\frac{\epsilon}{2\xi}\Bigr)^{k}  \nonumber  \\
&&\hspace{0.5cm}=\frac{e^{-|\xi|^{2}/4}}{(2\pi)^{1/4}}\sqrt{(2k+1)!}\sum_{r=0}^{k} ~_{2k+1-r}C_{r}~\frac{\xi^{2k+1-r}}{(2k+1-r)!}\Bigl(-\frac{\epsilon}{2\xi}\Bigr)^{r}  \nonumber  \\
&&\hspace{0.5cm}=\frac{e^{-|\xi|^{2}/4}}{(2\pi)^{1/4}}\sqrt{(2k+1)!}~\xi^{2k+1} \sum_{r=0}^{k} ~\frac{1}{ r! (2k+1-2r)!} \Bigl(-\frac{\epsilon}{2\xi^{2}}\Bigr)^{r}.
\label{eq:sakoda13}
\end{eqnarray}
Now  Hermite polynomials defined by
\begin{equation}
H_{n}(x)=(-1)^{n} e^{x^{2}/2} ~\frac{d^{n}}{d x^{n}}e^{-x^{2}/2}=\sum_{r=0}^{[n/2]}(-1)^{r} (2r-1)!! ~_{n}C_{2r}~x^{n-2r}
\label{eq:Hd}
\end{equation}
takes the following forms for $n={\rm even}$ or odd:
\begin{eqnarray}
&&H_{2k}(x)=x^{2k} \sum_{r=0}^{k} ~\frac{(2r-1)!!(2k)!}{(2r)!(2k-2r)!}\Bigl(-\frac{1}{x^{2}}\Bigr)^{r},  \nonumber \\
&&\hspace{1.5cm}=(2k)!x^{2k} \sum_{r=0}^{k}~\frac{1}{r!(2k-2r)!}\Bigl(-\frac{1}{2x^{2}}\Bigr)^{r}   \nonumber  \\
&&H_{2k+1}(x)=x^{2k+1} \sum_{r=0}^{k} ~\frac{(2r-1)!!(2k+1)!}{(2r)!(2k+1-2r)!}\Bigl(-\frac{1}{x^{2}}\Bigr)^{r}  \nonumber \\
&&\hspace{1.5cm}=(2k+1)!x^{2k+1} \sum_{r=0}^{k}~\frac{1}{r!(2k+1-2r)!}\Bigl(-\frac{1}{2x^{2}}\Bigr)^{r} ,\nonumber \\
\label{eq:sakoda14}
\end{eqnarray}
 where use is made
of the relation
\begin{equation}
(2r)!=(2r)(2r-1)(2r-2)(2r-3)\cdots\times 2\cdot 1=2^{r}r!(2r-1)!! ~~.
\end{equation}
Putting, in (\ref{eq:sakoda12}) and (\ref{eq:sakoda13}), $\epsilon=+, \xi=\xi_{r}, \epsilon/\xi^{2}=1/\xi_{r}^{2}$ or $\epsilon=-1, \xi=i\xi_{i}, \epsilon/\xi^{2}=1/\xi_{i}^{2}$ and using (\ref{eq:sakoda14}), we arrive at (\ref{eq:H1}) and (\ref{eq:H2}).

\subsection{Proofs of (\ref{eq:ortho3}) and (\ref{eq:transf3}) }

The whole arguments run in parallel with the case of Appendix {\bf 1}. We begin by  noticing the relation
\begin{equation}
e^{-\gamma ab} (\xi a^{\dagger}+\eta b^{\dagger}) e^{\gamma ab}=(\xi a^{\dagger}+\eta b^{\dagger})-\gamma (\xi b + \eta a),
\label{eq:sakoda15}
\end{equation}
we have
\begin{equation}
e^{-\gamma ab} e^{\xi a^{\dagger}+\eta b^{\dagger}}=e^{\xi a^{\dagger}+\eta b^{\dagger}-\gamma(\xi b+\eta a)} e^{-\gamma ab}.
\label{eq:sakoda16}
\end{equation}
Next, by putting $N=(a^{\dagger}a+b^{\dagger}b)$ we have
\begin{equation}
e^{-\beta N}(\xi a^{\dagger}+\eta b^{\dagger}-\gamma \xi b-\gamma \eta a)e^{\beta N}=e^{-\beta}(\xi a^{\dagger}+\eta b^{\dagger})-e^{\beta} \gamma(\xi b +\eta a),
\label{eq:sakoda17}
\end{equation}
so that
\begin{eqnarray}
e^{-\beta N}e^{-\gamma ab}e^{\xi a^{\dagger}+\eta b^{\dagger}}&=&e^{-\beta N}e^{\xi a^{\dagger}+\eta b^{\dagger}-\gamma (\xi b+\eta a)}e^{\beta N}e^{-\beta N}e^{-\gamma ab}  \nonumber  \\
&=&\exp\{e^{-\beta}(\xi a^{\dagger}+\eta b^{\dagger})-e^{\beta} \gamma (\xi b+\eta a)\}e^{-\beta N}e^{-\gamma ab}.
\label{eq:sakoda18}
\end{eqnarray}
Similarly, we have also
\begin{equation}
e^{\xi^{\prime *}a+\eta^{\prime *} b}e^{-\alpha a^{\dagger}b^{\dagger}}=e^{-\alpha a^{\dagger}b^{\dagger}} e^{\xi^{\prime *}a+\eta^{\prime *}b-\alpha(\xi^{\prime *}b^{\dagger}+\eta^{\prime *}a^{\dagger})}.
\label{eq:sakoda19}
\end{equation}
Using these relations we can then proceed as follows:\begin{eqnarray}
&&<0|e^{(\xi^{\prime *}a+\eta^{\prime *}b)}  e^{-\alpha  a^{\dagger}b^{\dagger}} e^{-\beta N} e^{-\gamma ab} e^{(\xi a^{\dagger}+\eta b^{\dagger})}|0>  \nonumber  \\
  &&~~~=<0|e^{-\alpha a^{\dagger}b^{\dagger}}e^{\xi^{\prime *}a+\eta^{\prime *}b-\alpha(\xi^{\prime *}b^{\dagger}+\eta^{\prime *}a^{\dagger})}e^{(\xi a^{\dagger}+\eta b^{\dagger})e^{-\beta}-\gamma(\xi b+\eta a)e^{\beta}} e^{-\beta N}e^{-\gamma ab}|0> \nonumber  \\
&&~~~=<0|e^{\xi^{\prime *}a+\eta^{\prime *}b-\alpha(\xi^{\prime *}b^{\dagger}+\eta^{\prime *}a^{\dagger})}e^{(\xi a^{\dagger}+\eta b^{\dagger})e^{-\beta}-\gamma (\xi b+\eta a)e^{\beta}} |0>e^{-\beta} \nonumber  \\
&&~~~=e^{-\alpha \xi^{\prime *}\eta^{\prime *} }e^{-\gamma \xi \eta} <0|e^{\xi^{\prime *}a}e^{\eta^{\prime *}b} e^{\xi e^{-\beta}a^{\dagger}}e^{\eta e^{\beta}b^{\dagger}}|0>e^{-\beta} \nonumber  \\
&&~~~=e^{-\alpha \xi^{\prime *}\eta^{\prime *} }e^{-\gamma \xi \eta} e^{\xi^{\prime *}\xi e^{-\beta}}e^{\eta^{\prime *}\eta e^{\beta}}e^{-\beta}.
\label{eq:sakoda2-1}
\end{eqnarray}

In a  way similar to the case of (\ref{eq:sakoda1}) we can show that
\begin{equation}
e^{\pm \lambda ab}e^{- \lambda a^{\dagger}b^{\dagger}}=e^{-\tilde{\alpha}_{\pm}( \lambda)a^{\dagger}b^{\dagger}}e^{-\tilde{\beta}_{\pm}( \lambda)N}e^{-\tilde{\gamma}_{\pm}( \lambda)ab},
\label{eq:sakoda2-2}
\end{equation}
where $-1<\lambda <1$ and 
\begin{equation}
\tilde{\alpha}_{\pm}(\lambda)=\tilde{\gamma}_{\pm}(\lambda)=\frac{\lambda}{1\pm \lambda^{2}},~~~\tilde{\beta}_{\pm}(\lambda)=\log (1\pm \lambda^{2}). \label{eq:sakoda2-5}
\end{equation}
Using (\ref{eq:sakoda2-2}) we can  write
\begin{eqnarray}
&& <0|e^{(\xi^{\prime *}a+\eta^{\prime *}b)} e^{\pm\lambda ab} e^{- \lambda a^{\dagger}b^{\dagger}}e^{(\xi a^{\dagger}+\eta b^{\dagger})}|0>   \nonumber  \\
&&~~~~~~~~=<0|e^{(\xi^{\prime *}a+\eta^{\prime *}b)}  e^{-\tilde{\alpha}_{\pm}   a^{\dagger}b^{\dagger}} e^{-\tilde{\beta}_{\pm}  N} e^{-\tilde{\gamma}_{\pm}  ab} e^{(\xi a^{\dagger}+\eta b^{\dagger})}|0> , 
\label{eq:sakoda2-6}
\end{eqnarray}
which owing to (\ref{eq:sakoda2-1}) becomes
\begin{equation}
=e^{-\tilde{\alpha}_{\pm}  \xi^{\prime *}\eta^{\prime *} }e^{-\tilde{\gamma}_{\pm}  \xi \eta} e^{\xi^{\prime *}\xi e^{-\tilde{\beta}_{\pm} }}e^{\eta^{\prime *}\eta e^{-\tilde{\beta}_{\pm} }}e^{-\tilde{\beta}_{\pm} }.
\label{eq:sakoda2-7}
\end{equation}

In the last expression we substitute $\tilde{\alpha}_{-}, \tilde{\beta}_{-}, \tilde{\gamma}_{-}$, given by  (\ref{eq:sakoda2-5}), $\eta=\epsilon \xi^{\ast}, \eta^{\prime}=\epsilon \xi^{\prime \ast}$ , and $\lambda=\pm (1- \kappa)$ according as $\epsilon =\pm$. Then the resulting expression leads, in the limit $\kappa \to 0$, to (\ref{eq:sakoda2-6}) when   use is made of the formula
\begin{equation}
\lim_{\kappa \to 0} \frac{1}{2\pi \kappa}\exp\Bigl(-\frac{x^{2}+y^{2}}{2\kappa}\Bigr)=\delta(x)\delta(y).
\label{eq:Dirac2}
\end{equation}

Similarly, in  (\ref{eq:sakoda2-7}) we substitute $\tilde{\alpha}_{+}, \tilde{\beta}_{+}, \tilde{\gamma}_{+}$ given by (\ref{eq:sakoda2-5}), $\eta=\xi^{\ast}, \eta^{\prime}=-\xi^{\prime \ast}$ and $\lambda \approx 1$, thereby obtaining the expression corresponding to (\ref{eq:sakoda2-7}) in the text.

\subsection{Proof of (\ref{eq:nm})}
The relation 
\begin{eqnarray}
|\xi,\eta>_{\epsilon}&\equiv &e^{-\epsilon a^{\dagger}b^{\dagger}}e^{\xi a^{\dagger}+\eta b^{\dagger}}|0>=e^{(-\epsilon b^{\dagger}+\xi)a^{\dagger}}e^{\eta b^[\dagger}|0> \nonumber  \\
&=&\sum_{\ell=0}^{\infty} \frac{(-\epsilon b^{\dagger}+\xi)^{\ell}}{\ell !}e^{\eta b^{\dagger}}(a^{\dagger})^{\ell}|0>
\end{eqnarray}
immediately leads us to
\begin{eqnarray}
&&<n,m|\xi,\eta>_{\epsilon} \nonumber \\
&&\hspace{0.5cm}=\frac{1}{\sqrt{n!m!}}<0|a^{n}b^{m}\sum_{\ell=0}^{\infty} \frac{(-\epsilon b^{\dagger}+\xi)^{\ell}}{\ell !}e^{\eta b^{\dagger}}(a^{\dagger})^{\ell}|0>  \nonumber  \\
&&\hspace{0.5cm}=\frac{1}{\sqrt{n!m!}}<0|b^{m}e^{\eta b^{\dagger}}(-\epsilon b^{\dagger}+\xi)^{n}|0>  \nonumber  \\
&&\hspace{0.5cm}=\frac{1}{\sqrt{n!m!}}<0|(b+\eta)^{m}(-\epsilon b^{\dagger}+\xi)^{n}|0>  \nonumber  \\
&&\hspace{0.5cm}=\frac{1}{\sqrt{n!m!}}\sum_{r=0}^{m} \sum_{s=0}^{n} \frac{m!}{r!(m-r)!}\frac{n!}{s!(n-s)!} \eta^{m-r} \xi^{n-s}(-\epsilon)^{s} <0|b^{r} (b^{\dagger})^{s}|0> \nonumber  \\
&&\hspace{0.5cm}=\frac{1}{\sqrt{n!m!}}\sum_{r=0}^{m} \sum_{s=0}^{n} \frac{m!}{r!(m-r)!}\frac{n!}{s!(n-s)!} \eta^{m-r} \xi^{n-s}(-\epsilon)^{s} r! \delta_{rs},
\end{eqnarray}
which for $\eta=\epsilon \xi^{\ast}$ gives (\ref{eq:nm}) when (\ref{eq:Laguerre1}) is taken into account.

\subsection{Proof of (\ref{eq:complete3})}
By using (\ref{eq:state3-4}) and putting $\xi=x+iy=r e^{i\theta}$ we can proceed as follows:

\begin{eqnarray}
&& \int dx dy~||\xi,\epsilon \xi^{\ast}><\xi,\epsilon \xi^{\ast}|  \nonumber  \\&&=\frac{1}{\pi}\int dx dy e^{-|\xi|^{2}/2} \Bigl\{ \sum_{n=0}^{\infty} \sum_{m=0}^{n}~\sqrt{\frac{m!}{n!}}~\xi^{n-m}~(-\epsilon)^{m}~L_{m}^{n-m}(|\xi|^{2})~|n,m>  \nonumber  \\
&&\hspace{3.5cm}+\sum_{m=1}^{\infty} \sum_{n=0}^{m-1}~\sqrt{\frac{n!}{m!}}~(\epsilon \xi^{\ast})^{m-n}~(-\epsilon)^{n}~L_{n}^{m-n}(|\xi|^{2})~|n,m> \Bigr\}  \nonumber  \\
&&\hspace{3cm}\times  \Bigl\{ \sum_{n^{\prime}=0}^{\infty} \sum_{m^{\prime}=0}^{n^{\prime}}~\sqrt{\frac{m^{\prime}!}{n^{\prime}!}}~(\xi^{\ast})^{n^{\prime}-m^{\prime}}~(-\epsilon)^{m^{\prime}}~L_{m^{\prime}}^{n^{\prime}-m^{\prime}}(|\xi|^{2})<n^{\prime},m^{\prime}|  \nonumber  \\
&&\hspace{3.5cm}+\sum_{m^{\prime}=1}^{\infty} \sum_{n^{\prime}=0}^{m^{\prime}-1}~\sqrt{\frac{n^{\prime}!}{m^{\prime}!}}~(\epsilon \xi)^{m^{\prime}-n^{\prime}}~(-\epsilon)^{n^{\prime}}~L_{n^{\prime}}^{m^{\prime}-n^{\prime}}(|\xi|^{2})<n^{\prime},m^{\prime}| \Bigr\}  \nonumber  \\
&&~=\frac{1}{\pi}\int dx dy e^{-|\xi|^{2}/2} \Bigl\{ \sum_{n=0}^{\infty} \sum_{m=0}^{n}\sum_{n^{\prime}=0}^{\infty}\sum_{m^{\prime}=0}^{n^{\prime}}~\sqrt{\frac{m!m^{\prime}!}{n!n^{\prime}!}}~\xi^{n-m}(\xi^{\ast})^{n^{\prime}-m^{\prime}}~(-\epsilon)^{m+m^{\prime}}~  \nonumber  \\
&&\hspace{6cm} \times L_{m}^{n-m}(|\xi|^{2})L_{m^{\prime}}^{n^{\prime}-m^{\prime}}(|\xi|^{2})~|n,m><n^{\prime},m^{\prime}|   \nonumber  \\
&&\hspace{3cm}+\sum_{n=0}^{\infty} \sum_{m=0}^{n}\sum_{m^{\prime}=1}^{\infty}\sum_{n^{\prime}=0}^{m^{\prime}-1}~\sqrt{\frac{m!n^{\prime}!}{n!m^{\prime}!}}~\xi^{n-m}(\epsilon\xi)^{m^{\prime}-n^{\prime}}~(-\epsilon)^{m+n^{\prime}}~  \nonumber  \\
&&\hspace{5cm} \times L_{m}^{n-m}(|\xi|^{2})L_{n^{\prime}}^{m^{\prime}-n^{\prime}}(|\xi|^{2})~|n,m><n^{\prime},m^{\prime}|   \nonumber  \\
&&\hspace{3cm}+\sum_{m=1}^{\infty} \sum_{n=0}^{m-1}\sum_{n^{\prime}=0}^{\infty}\sum_{m^{\prime}=0}^{n^{\prime}}~\sqrt{\frac{n!m^{\prime}!}{m!n^{\prime}!}}~(\epsilon\xi^{\ast})^{m-n}(\xi^{\ast})^{n^{\prime}-m^{\prime}}~(-\epsilon)^{n+m^{\prime}}~  \nonumber  \\
&&\hspace{5cm} \times L_{n}^{m-n}(|\xi|^{2})L_{m^{\prime}}^{n^{\prime}-m^{\prime}}(|\xi|^{2})~|n,m><n^{\prime},m^{\prime}|   \nonumber  \\
&&\hspace{3cm}+\sum_{m=1}^{\infty} \sum_{n=0}^{m-1}\sum_{m^{\prime}=1}^{\infty}\sum_{n^{\prime}=0}^{m^{\prime}-1}~\sqrt{\frac{n!n^{\prime}!}{m!m^{\prime}!}}~(\epsilon \xi^{\ast})^{m-n}(\epsilon\xi)^{m^{\prime}-n^{\prime}}~(-\epsilon)^{n+n^{\prime}}~  \nonumber  \\
&&\hspace{5cm} \times L_{n}^{m-n}(|\xi|^{2})L_{n^{\prime}}^{m^{\prime}-n^{\prime}}(|\xi|^{2})~|n,m><n^{\prime},m^{\prime}| \Bigr\}.
\label{eq:sakoda3-0}
\end{eqnarray}

The first term in the above exression can now be written as
\begin{eqnarray}
&&\sum_{n=0}^{\infty}\sum_{m=0}^{n}\sum_{n^{\prime}}^{\infty}\sum_{m^{\prime}}^{n^{\prime}}\sqrt{\frac{m!m^{\prime}!}{n!n^{\prime}!}}(-\epsilon)^{m+m^{\prime}} |n,m><n^{\prime},m^{\prime}| \nonumber  \\
&&\hspace{0.5cm} \times \Bigl\{\frac{1}{2\pi}\int d(r^{2}) d\theta e^{-r^{2}}r^{n-m+n^{\prime}-m^{\prime}}~e^{i\theta (n-m-n^{\prime}+m^{\prime})}~L_{m}^{n-m}(r^{2})~L_{m^{\prime}}^{n^{\prime}-m^{\prime}}(r^{2})\Bigr\}   \nonumber  \\
&&\hspace{0.5cm} =\sum_{n=0}^{\infty}\sum_{m=0}^{n}\sum_{n^{\prime}}^{\infty}\sum_{m^{\prime}}^{n^{\prime}}\sqrt{\frac{m!m^{\prime}!}{n!n^{\prime}!}}(-\epsilon)^{m+m^{\prime}} |n,m><n^{\prime},m^{\prime}| \nonumber  \\
&&\hspace{2cm} \times \Bigl\{\int d(r^{2})  e^{-r^{2}}(r^{2})^{n-m}~L_{m}^{n-m}(r^{2})~L_{m^{\prime}}^{n-m}(r^{2})\Bigr\}\delta_{n-m,n^{\prime}-m^{\prime}}   \nonumber  \\
&&\hspace{0.5cm} =\sum_{n=0}^{\infty}\sum_{m=0}^{n} \frac{m!}{n!}\frac{\Gamma(n+1)}{m!} |n,m><n,m| =\sum_{n=0}^{\infty}\sum_{m=0}^{n}|n,m><n,m|.
\label{eq:sakoda3-1}
\end{eqnarray}
Similarly the fourth term therein becomes
\begin{equation}
\sum_{m=1}^{\infty}\sum_{n=0}^{m-1}|n,m><n,m|. 
\end{equation}
On the other hand the $\theta$-integration  $\int d \theta e^{i(n-m+m^{\prime}-n^{\prime})}$ in the second term vanishes since $n-m+m^{\prime}-n^{\prime} \ne 0$, so
that this term gives no contribution to (\ref{eq:sakoda3-0}). Likewise for the third term. 

Collecting the above results, we are led to
\begin{equation}
\int dx dy ~|\xi,\epsilon \xi^{\ast}><\xi, \epsilon \xi^{\ast}|=\sum_{n=0}^{\infty}\sum_{m=0}^{n}|n,m><n,m|+\sum_{m=1}^{\infty}\sum_{n=0}^{m-1}|n,m><n,m|={\rm I}
\end{equation}
that is, (\ref{eq:complete3}).

\end{document}